\documentclass[10pt,a4paper,english,pra,twocolumn]{revtex4-1}
\usepackage[LGR,T1]{fontenc}
\usepackage{textcomp}
\usepackage[latin9]{inputenc}
\usepackage{color}
\definecolor{page_backgroundcolor}{rgb}{1, 1, 1}
\pagecolor{page_backgroundcolor}
\usepackage{array}
\usepackage{booktabs}
\usepackage{multirow}
\usepackage{amsmath}
\usepackage{amssymb}
\usepackage{graphicx}

\makeatletter

\newcommand*\LyXZeroWidthSpace{\hspace{0pt}}
\DeclareRobustCommand{\greektext}{%
  \fontencoding{LGR}\selectfont\def\encodingdefault{LGR}}
\DeclareRobustCommand{\textgreek}[1]{\leavevmode{\greektext #1}}

\providecommand{\tabularnewline}{\\}

\usepackage{array}     
\usepackage{booktabs}  
\usepackage{multirow}  
\usepackage{multirow}
\usepackage{diagbox}
\usepackage{booktabs}
\usepackage[colorlinks = true,
            linkcolor = red,
            urlcolor  = blue,
            citecolor = blue,
            anchorcolor = blue]{hyperref}
\begingroup

\footnotetext{These authors contributed equally to this work.}
\endgroup

\makeatother

\usepackage{babel}
\begin{document}
\title{Multiphoton heralding generates large-amplitude squeezed Schr\"odinger
cat states and parity\nobreakdash-selective Fock superpositions from
squeezed vacuum via an OPA}
\author{Yusuf Turek}
\email{Corresponding author: yusufu1984@hotmail.com}

\author{Ming-Yan Sun\textsuperscript{\S}}
\author{Xiao-Xi Yao\textsuperscript{\S}}
\affiliation{School of Physics,Liaoning University,Shenyang,Liaoning 110036,China}
\date{\today}
\begin{abstract}
We propose a multiphoton heralding scheme using an optical parametric
amplifier (OPA) that converts squeezed vacuum into two families of
non\nobreakdash-Gaussian states: large\nobreakdash-amplitude squeezed
Schr\"odinger cat states and low\nobreakdash-order parity\nobreakdash-selective
Fock superpositions. By injecting $m$ photons into the idler port
and detecting $n$ photons at the output, effective high\nobreakdash-order
photon subtraction is realized in a single OPA device. The heralded
states exhibit strong Wigner negativity and high phase\nobreakdash-space
complexity. Remarkably, under photon loss, the complexity remains
substantial even after negativity vanishes, indicating a loss\nobreakdash-resilient
quantum resource. These states also surpass the Heisenberg limit in
phase estimation. Our protocol establishes the OPA as a versatile
platform for generating non\nobreakdash-Gaussian states, with promising
applications in loss\nobreakdash-resilient quantum metrology and
fault\nobreakdash-tolerant quantum information processing.
\end{abstract}
\maketitle

\section{\label{sec:1}Introduction}

Gaussian states such as squeezed vacua, coherent states, and thermal
states are the workhorses of continuous-variable (CV) quantum information,
enabling protocols such as quantum key distribution and boson sampling
\citep{RevModPhys.77.513}. However, they form a classical submanifold
in phase space and are insufficient for universal CV quantum computation
or fault-tolerant quantum error correction without additional non-Gaussian
resources \citep{PRXQuantum.1.020305,lachman2022quantum,PRXQuantum.2.030204}.
Non-Gaussian states, particularly Schr\"odinger cat (SC) states \citep{PRXQuantum.3.010301,PhysRevLett.132.230602}
, photon-added states \citep{89} , and Fock-state superpositions
\citep{PhysRevLett.118.223604,PhysRevA.110.042421}, overcome this
limitation by enabling universal gate sets \citep{PhysRevA.93.022301,PRXQuantum.2.010327,PRXQuantum.6.010311},
bosonic code error correction \citep{PRXQuantum.2.020101,2024Q,PhysRevA.55.900,PhysRevX.8.021054},
quantum computation \citep{PhysRevA.68.042319}, and quantum metrology\citep{2015SR,PhysRevA.91.013808,PhysRevA.88.013838,PhysRevA.93.033859}
beyond the standard quantum limit. Among these, parity-selective Fock
superpositions including large-amplitude squeezed cat states are essential
resources for correcting photon loss and dephasing errors \citep{PhysRevX.6.031006,hr5f-lvy7}
and for encoding Gottesman-Kitaev-Preskill (GKP) qubits \citep{R2023}.

A standard approach to generating non-Gaussian states is conditional
measurement on a beam splitter \citep{PhysRevA.55.3184,2006,PhysRevLett.101.233605,PhysRevA.82.031802}.
In that protocol, $k$-photon subtraction can realize a certain type
of SC state with amplitude at most $\alpha\sim\sqrt{k}$ with high-fidelity.
For example, single-photon subtraction from a squeezed vacuum (SV),
heralded by single\nobreakdash-photon detection, produces a small\nobreakdash-amplitude
( $\alpha\le1$) odd cat state with fidelity exceeding $0.99$ \citep{2007S,ourjoumtsev2007generation,Wakui:07,Takase:22}.
For fault-tolerant bosonic error correction, however, SC states with
amplitude $\alpha\ge2$ are required to ensure near-orthogonality
between the two coherent components ( $e^{-2\alpha^{2}}\ll1$). Achieving
such amplitudes via conventional photon subtraction demands detecting
at least four photons, leading to a success probability scaling as
$R^{k}$ with typical beam-splitter reflectivity $R\sim0.02-0.05$
which leads to extremely low generation rates \citep{RN20}. Generalized
photon subtraction (GPS) \citep{PhysRevA.103.013710,PhysRevA.110.033717},
postseclected weak measurement technique \citep{npj2026,PhysRevA.105.022608,QST2026},
distillation of squeezing \citep{PhysRevA.111.043704,PhysRevLett.129.273604,PhysRevLett.112.070402}
and quantum optical catalysis (QOC) \citep{PhysRevLett.88.250401,ktc9-9rjb}
offer improvements but remain tailored to specific target states and
lack flexibility.

A promising alternative is to replace the beam splitter with an optical
parametric amplifier (OPA). Unlike a beam splitter, an OPA simultaneously
enables photon addition and subtraction, and its gain provides a tunable
knob for state engineering. In 2019, Shringarpure and Franson showed
that an OPA can prepare photon-added states without actual photon
addition \citep{PhysRevA.100.043802}. Our group later extended this
to SV inputs with single-photon injection and detection, realizing
effective two-photon subtraction and a low-order even SC state \citep{rkzg-sdxn}.
However, these schemes were restricted to single-photon idler inputs
and single-photon detection.

In this work, we answer the natural question: Can multiphoton input
and detection in an OPA realize effective $k\ge3$ photon subtraction?
We provide an affirmative answer. By systematically varying the number
of idler input photons $m$ and detected photons $n$ (with SV in
the signal port), we show that the OPA can effectively implement $\kappa\ge3$
photon subtraction for $k=m+n$. Crucially, not every ($m$, $n$)
pair satisfying $k=m+n$ yields a well approximation to a certain
type of SC state. Only specific configurations, identified by systematically
optimizing the OPA gain $g$ and the input squeezing $r$, produce
high\nobreakdash-fidelity target states. For example, ($1$, $2$)
and ($4$, $1$) generate squeezed odd Shrodinger cat (SOSC) states
with amplitudes $\alpha\approx\sqrt{3}$ and $\alpha\approx\sqrt{5}$,
respectively, with fidelities exceeding $0.99$. These amplitudes
are comparable to those achieved by four- or five-photon subtraction
in conventional schemes, but with dramatically higher success probabilities
(e.g., $P_{trial}\approx1.15\%$ for ($1$, $2$) vs $R^{4}\sim10^{-4}$
for BS subtraction). While our previous study {[}24{]} focused exclusively
on the ($1$, $1$ ) configuration, the present work extends the analysis
to multiphoton input and detection, revealing a general parity selection
rule and identifying optimal ($m$, $n$) pairs for generating high\nobreakdash-fidelity
SC states and higher\nobreakdash-order Fock superpositions. Beyond
state engineering, we further demonstrate the advantages of these
states in phase estimation and their resilience against photon loss
--- aspects not considered before. 

In addition to Wigner negativity, we introduce a complementary metric
--- phase-space complexity $\mathcal{C}$ --- to capture structural
richness that can persist even when negativity vanishes under photon
loss. Detailed definitions and analysis are given in Sec. \ref{sec:3}.
Our protocol thus transforms the OPA from a specialized source for
low-order SC states into a versatile, integrated, and reconfigurable
platform for a broad class of high-photon-number non-Gaussian states.

The paper is organized as follows. After introducing the general heralding
protocol in Sect. \ref{sec:2}, we analyze the properties of the heralded
states from a SV input, including state engineering (Secs. \ref{sec:3}
and \ref{sec:4}) and phase estimation (Sec. \ref{sec:5}). Section
\ref{sec:6}, discusses practical limitations of our protocol. Concluding
remarks are provided in Sec. \ref{sec:7}. Numerical simulations of
this work are done using QuTiP \citep{JOHANSSON20131234}.

\section{\label{sec:2} The non-Gaussian state source}

\begin{figure}
\includegraphics[width=8cm]{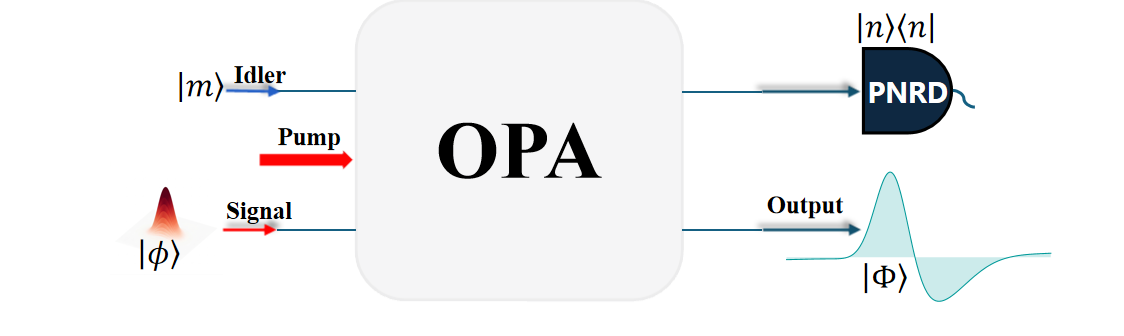}

\caption{\label{fig:1}Schematic of the non-Gaussian source using an optical
parametric amplifier (OPA). Specific states $\vert\phi\rangle$--
such as coherent, squeezed vacuum (SV) and small-amplitude Schr\"odinger
cat (SC) states - are injected into the signal mode, while $m$-photon
is input into the idler mode of the OPA. $m$-photon detection at
the idler output port by the photon-number-resolving detection (PNRD)
heralds the generation of a non-Gaussian quantum state $\vert\Phi\rangle$
at the signal output. The characteristics of the output state can
be controlled by adjusting squeezing parameter $r$, input and detected
photon numbers $n$ and $m$, and the gain $g$ of the amplifier,
respectively.}
\end{figure}

Figure \ref{fig:1} illustrates a generic method for generating non-Gaussian
states from Gaussian and non-Gaussian inputs. In our protocol, we
consider general formalism of state engineering issues by taking Gaussian
and non-Gaussian inputs into two input ports of an OPA. As shown in
Fig. \ref{fig:1}, $m$ photon ($\vert m\rangle$) and arbitrary state
$\vert\phi\rangle$ are plugged into the idler and signal modes of
the OPA, then OPA produces two correlated output modes. The photon-number-resolving
detection (PNRD) on the idler port naturally heralds a non-Gaussian
state in the signal output port. Our scheme supports multiphoton heralding,
so that a much broader class of non-Gaussian states can generated
with various input states. 

The OPA in our setup can be described by the two-mode squeezing unitary
as 
\begin{align}
S(\tau) & =\exp\left[\tau^{\ast}ab-\tau a^{\dagger}b^{\dagger}\right]\nonumber \\
 & =\frac{1}{g}e^{-Ga^{\dagger}b^{\dagger}}g^{-(a^{\dagger}a+b^{\dagger}b)}e^{Gab}\label{eq:1}
\end{align}
 where $\tau=\varrho e^{i\delta}$and $g=\cosh\varrho$ denote the
non-linear and gain parameters of the OPA, $G=\tanh\tau=\sqrt{g^{2}-1}/g$,
and $a$ ($a^{\dagger}$) $b$ ($b^{\dagger}$) are the annihilation
(creation) operators of signal and idler modes, respectively. The
PNRD on the idler output mode can modelled by the projection operator
\begin{equation}
\Pi_{n}=\vert n\rangle\langle n\vert\label{eq:2}
\end{equation}
which corresponding to the detection of $n$ photons.

For our protocol, the unnormalised signal output state can be written
as 
\begin{align}
\vert\psi\rangle_{m,n} & =\langle n\vert\left(\mathbb{I}\otimes\Pi_{n}\right)S(\tau)\left(\vert\phi\rangle\otimes\vert m\rangle\right)\nonumber \\
 & =\sum^{m}_{k=0}H(k,m,n)(a^{\dagger})^{n-m+k}g^{-a^{\dagger}a}a^{k}\vert\phi\rangle,\label{eq:3}
\end{align}
 where 
\begin{equation}
H(k,m,n)=\frac{G^{n-m+2k}}{g^{m-k+1}}\frac{(-1)^{n-m+k}\sqrt{n!m!}}{k!(n-m+k)!(m-k)!}\label{eq:4}
\end{equation}
 and the success probability is given by $P_{m,n}(n,m,r,g)={}_{n,m}\langle\psi\vert\psi\rangle_{m,n}$.
The specific forms of $\vert\psi\rangle_{m,n}$ depends on initial
input signal state $\vert\phi\rangle$. In this work, the input signal
state to the OPA is primarily the SV state, which yields the non-Gaussian
heralded states that are the main focus of this paper. For completeness,
we also provide the definitions of the coherent state and the SC state.
The coherent state serves as a useful benchmark, while the SC state
will be used as a target state in our fidelity analysis.

1. Coherent state $\vert\alpha\rangle$
\begin{align}
\vert\alpha\rangle & =D(\alpha)\vert0\rangle\nonumber \\
 & =e^{-\frac{\vert\alpha\vert^{2}}{2}}\sum^{\infty}_{n=0}\frac{\alpha^{n}}{\sqrt{n!}}\vert n\rangle\label{eq:4-1}
\end{align}
 where $D(\alpha)=e^{(\alpha^{\ast}a-\alpha a^{\dagger})}$ is the
displacement operator and for simplicity we take $\alpha$ to be real. 

2. SV state $\vert\xi\rangle$ 
\begin{equation}
\vert\xi\rangle=S(z)\vert0\rangle\label{eq:SV}
\end{equation}
 where $S(z)=exp[\frac{1}{2}(z^{\ast}a^{2}-za^{\dagger2})]$ is the
single-mode squeezing operator with a complex squeezing parameter
$z=re^{i\theta}$. In the Fock basis,

\begin{align}
\vert\xi\rangle & =\frac{1}{\sqrt{\cosh r}}\sum^{\infty}_{n-0}\frac{\sqrt{(2n)!}}{n!2^{n}}(-\xi)^{n}\vert2n\rangle,\label{eq:4-2}
\end{align}
where $\xi=e^{i\theta}\tanh r$ and $\cosh r=(1-\vert\xi\vert^{2})^{1/4}$.
The squeezing level in decibels (dB) is given by $-20\log_{10}(e^{-r})$. 

3. SC state $\vert Cat\rangle_{\theta},\alpha$
\begin{equation}
\vert Cat\rangle_{\theta,\alpha}=N^{-1/2}_{\theta}\left(\vert\alpha\rangle+e^{i\theta}\vert-\alpha\rangle\right),\label{eq:cat}
\end{equation}
 where the normalization constant is $N_{\theta}=2\left(1+e^{-2\alpha^{2}}\cos\theta\right)$.
The cases $\theta=0$ and $\theta=\pi$ correspond to even and odd
SC states, respectively. The SC state is a fundamental state in CV
quantum information processing, as it can be seen as a rough approximation
to the GKP qubit \citep{PhysRevA.64.012310}.

In the recent work \citep{erkilic2025} Erlick et al. investigated
the non-Gaussian state generation proposal with the same setup with
ours but only considered vacuum idler input, i.e., $m=0$. For that
proposal the general form of the output state is given as (unnormalized)
\begin{align}
\vert\psi\rangle_{0,n} & =\frac{(-G)^{n}}{g}\frac{1}{\sqrt{n!}}(a^{\dagger})^{n}g^{-a^{\dagger}a}\vert\phi\rangle.\label{eq:3-1}
\end{align}
 Since $g^{-a^{\dagger}a}$ can only modify the amplitude of a given
state $\vert\phi\rangle\in\{\vert\alpha\rangle,\vert\xi\rangle,\vert Cat\rangle_{\theta},_{\alpha}\}$,
the $n$-photon detection event in idler output heralds $n$-photon
addition to the corresponding input state but with attenuated amplitudes,
i.e., $\vert\psi\rangle_{0,n}\propto\{(a^{\dagger})^{n}\vert\frac{\alpha}{g}\rangle,(a^{\dagger})^{n}\vert\frac{\xi}{g^{2}}\rangle,(a^{\dagger})^{n}\vert Cat\rangle_{\theta},_{\alpha/g}\}$
\citep{erkilic2025}. In this process the OPA takes the role of photon
addition to the given state. The special case of the above scheme
is that if one consider the photon vacuum state $\vert0\rangle$ zero
photon detection $\Pi_{0}$ in idler's mode (i.e., $m=n=0$), then
the OPA just takes the role of noiseless attenuation to the given
input signal state \citep{PhysRevA.96.042307}. 

In previous works, only $n=m=1$ case has been studied for coherent
state \citep{PhysRevA.100.043802}, SV and SC states \citep{rkzg-sdxn},
and confirmed the usefulness of OPA in non-Gaussian state generation.
As an extension of previous studies, in current work,  we mainly delve
the multiphoton detection ($n\ge2$) events for single photon $\vert1\rangle$
($m=1$) and $\vert m\rangle$ ($m\ge2$) idler input cases. 

\section{\label{sec:3}General description of SV input }

In the following, we concentrate on the SV input. Substituting $\vert\phi\rangle=\vert\xi\rangle$
into Eq. \ref{eq:3} and using the properties of the squeezing operator,
we obtain the unnormalized heralded state 
\begin{align}
\vert\Psi\rangle_{sv,m,n} & =\sum^{m}_{k=0}H(k,m,n)(a^{\dagger})^{n-m+k}g^{-a^{\dagger}a}a^{k}\vert\xi\rangle\nonumber \\
 & =\alpha(\xi,g)\sum^{m}_{k=0}H(k,m,n)g^{k}(a^{\dagger})^{n-m}(a^{\dagger}a)^{k}\vert\frac{\xi}{g^{2}}\rangle,\label{eq:9-2}
\end{align}
 where the factor $\alpha(\xi,g)$ is given by 
\begin{equation}
\alpha(\xi,g)=\frac{(1-\vert\xi\vert^{2})^{1/4}}{(1-\vert\frac{\xi}{g^{2}}\vert^{2})^{1/4}}.\label{eq:9}
\end{equation}
The input SV has even parity. The idler input contains $m$ photons
(parity $(-1)^{m}$), and the idler output is projected onto $n$
photons (parity $(-1)^{n}$). Because the OPA preserves total parity,
the parity of the heralded signal state must be $(-1)^{m+n}$. Consequently,
the output state $\vert\Psi\rangle_{sv,m,n}$ contains only even Fock
components when $m+n$ is even, and only odd Fock components when
$m+n$ is odd. This rule is general and holds for any ($m$, $n$). 

After applying the transformation that relates the squeezed state
$\vert\xi\rangle$, the state for $m=1$ and arbitrary $n$ cases
can be expressed as a sum of two dominant terms: 
\begin{align}
\vert\Psi\rangle_{sv,1,n} & =\nonumber \\
 & (-G)^{n-1}\frac{\alpha(\xi,g)}{g^{2}}\sqrt{\frac{n}{(n-1)!}}\sqrt{M_{n-1}(\vert\frac{\xi}{g^{2}}\vert)}\vert\frac{\xi}{g^{2}},n-1\rangle\nonumber \\
 & +\left(-G\right)^{n+1}\xi\frac{\alpha(\xi,g)}{g^{2}\sqrt{n!}}\sqrt{M_{n+1}(\vert\frac{\xi}{g^{2}}\vert)}\vert\frac{\xi}{g^{2}},n+1\rangle,\label{eq:9-1}
\end{align}
 where we have defined $n$-photon added squeezed state 
\begin{align}
\vert\xi,n\rangle & =\frac{(a^{\dagger})^{n}}{\sqrt{M_{n}(\vert\xi\vert)}}\vert\xi\rangle\nonumber \\
 & =\frac{(1-\vert\xi\vert^{2})^{1/4}}{\sqrt{M_{n}(\vert\xi\vert)}}\sum^{\infty}_{k=0}(-1)^{k}\left(\frac{\xi}{2}\right)^{k}\frac{\sqrt{(2k+n)!}}{k!}\vert2k+n\rangle.\label{eq:7-1}
\end{align}
 Here the normalization factor $M_{n}(\vert\xi\vert)$ is expressed
as: 
\begin{equation}
M_{n}(\vert\frac{\xi}{g^{2}}\vert)=n!\left(1-\vert\xi\vert^{2}\right)^{-n/2}P_{n}\left(\left(1-\vert\xi\vert^{2}\right)^{-1/2}\right),\label{eq:8-1}
\end{equation}
where $P_{n}(x)=\frac{1}{2^{n}n!}\frac{d^{n}}{dx^{n}}\left((x^{2}-1)^{n}\right)$is
the Legendre Polynomials. 

Due to the input SV contains only even Fock components, and the operator
$(a^{\dagger})^{n-1+k}g^{-a^{\dagger}a}a^{k}$ changes the parity
by $n-1+2k$ (mode $2$), the final state $\vert\Psi\rangle_{sv,1,n}$
contains only Fock states with parity opposite to that of $n$ (i.e.,
even $n$ yields odd Fock states, and odd $n$ yields even Fock states).
More explicitly, for odd $n$, the state can be written as 
\begin{equation}
\vert\Psi\rangle_{sv,1,n\in odd}\propto\sum^{\frac{n-1}{2}}_{k=0}c^{\prime}_{2k}\vert2k\rangle+\sum^{\frac{n+1}{2}}_{k=0}c^{\prime\prime}_{2k}\vert2k\rangle,
\end{equation}
 and for even $n$, the state contains only odd Fock components:
\begin{equation}
\vert\Psi\rangle_{sv,1,n\in even}\propto\sum^{\frac{n-1}{2}}_{k=0}c^{\prime}_{2k+1}\vert2k+1\rangle+\sum^{\frac{n}{2}}_{k=0}c^{\prime\prime}_{2k}\vert2k+1\rangle,
\end{equation}
The coefficients $c^{\prime}_{k}$ and $c^{\prime\prime}_{k}$ depend
on $r,g$ and $n$, and can be obtained from the Eq. (\ref{eq:9-1}).
Their explicit forms for $n=2,3,4$ are given in following subsections. 

As we can see, the signal output state $\vert\Psi\rangle_{sv,1,n}$
is a superposition of $(n-1)$-photon and $(n+1)$-photon added attenuated
squeezed states. And if the PNRD successfully detect multi-photon
($n\ge2$) then one can obtain the state $\vert\Psi\rangle_{sv,1,n}$
with success probability $P_{sv,1,n}(r,g,n)={}_{sv,1,n}\langle\Psi\vert\Psi\rangle_{sv,1,n}$. 
\begin{flushleft}
\begin{figure*}
\begin{centering}
\includegraphics[width=16cm,totalheight=16cm]{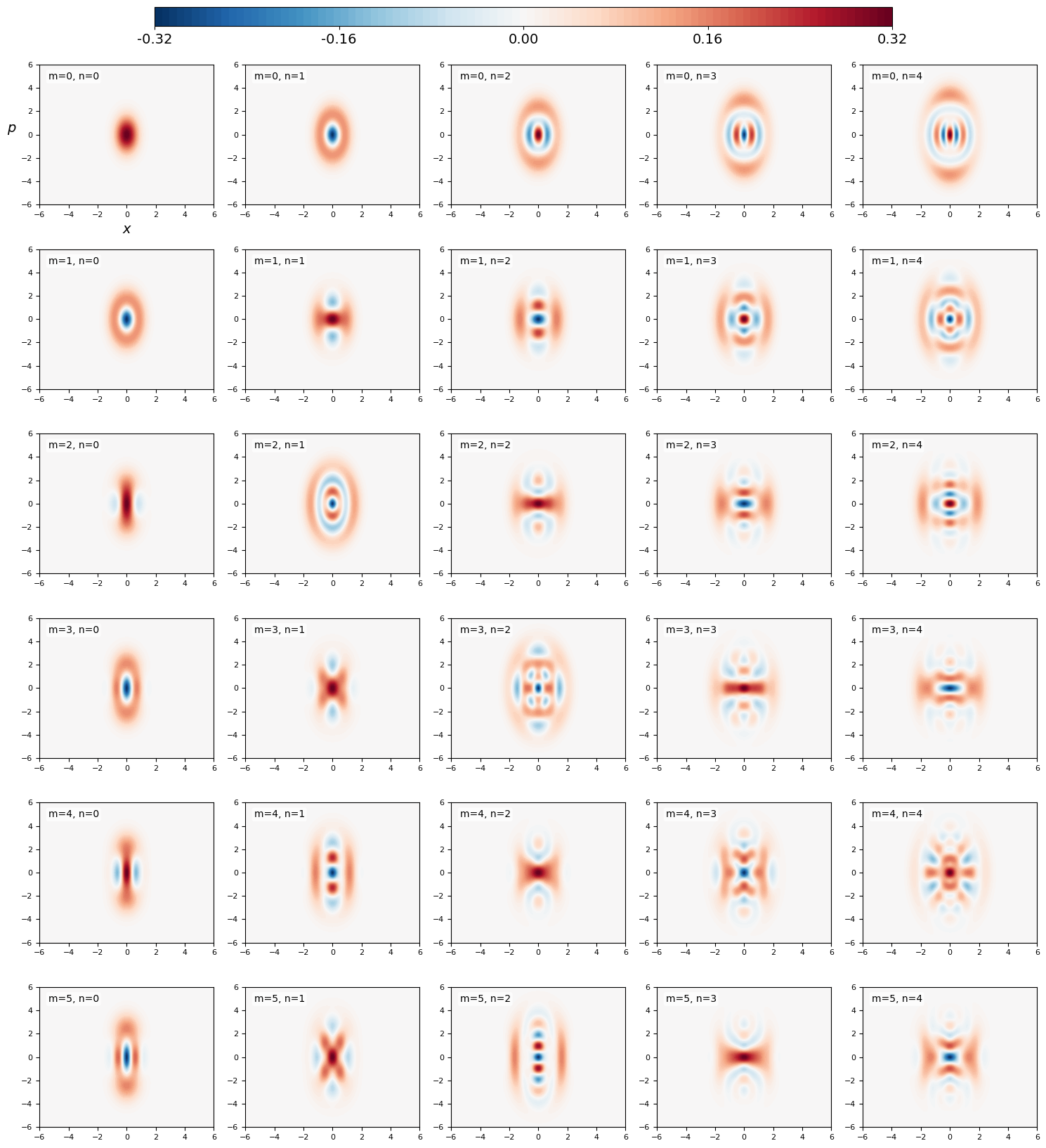}
\par\end{centering}
\caption{\label{fig:2}Wigner functions of the signal output state $\vert\Psi\rangle_{sv,m,n}$
for $m=0,1,2,3,4,5$ and $n=0,1,2,3,4$ cases. Here we take $r=1.0$
($\approx8.69\ dB$), $\theta=\varphi=0$ and $g=1.5$. }
\end{figure*}
\par\end{flushleft}

\begin{figure}
\includegraphics[width=8cm]{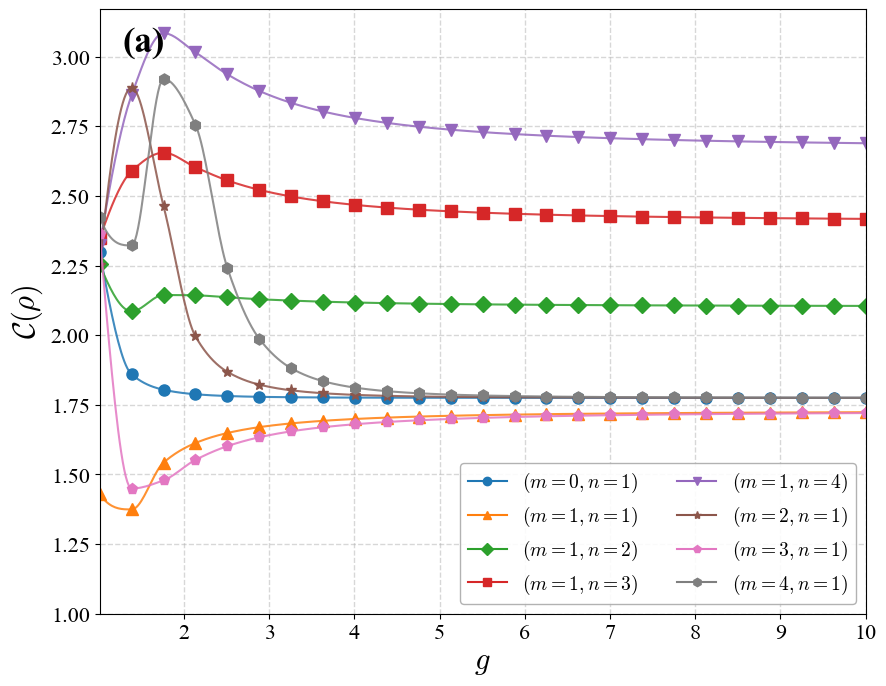}

\includegraphics[width=8cm]{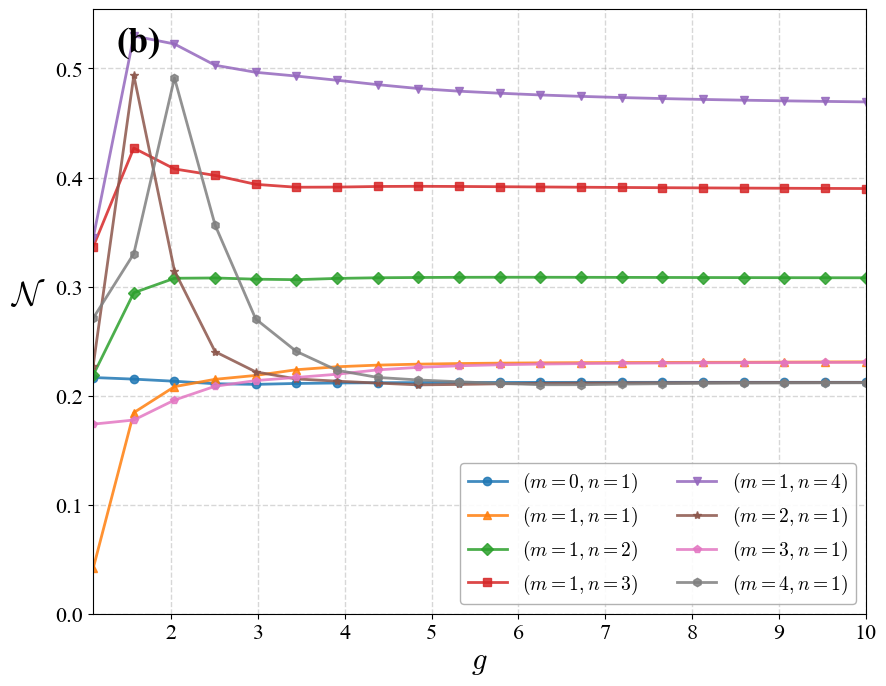}

\caption{\label{fig:4} Complexity $\mathcal{C}$ (a) and Wigner negativity
volume $\mathcal{N}$ (b) as a function of OPA gain $g$ with $r=1.0$.
The other parameters are the same as in Fig. \ref{fig:2}.}
\end{figure}

The Wigner functions of signal output state $\vert\Psi\rangle_{sv,m,n}$
for $m=0,1,2,3,4,5$ and $n=0,1,2,3,4$ corresponding to the SV input
case for fixed squeezing parameter $r=1.0$ ($\approx8.69$dB) and
OPA gain $g=1.5$ are showed in Fig. \ref{fig:2}. As indicated in
colorful figures  the non-classicality (negative region of Wigner
function) of signal output states are increasing with increase the
photon detection {[}also see the Fig. \ref{fig:4} (b) for this phenomena{]}.
From these plots of phase space distribution we can speculate that
similar as $n=1$ case (studied in Ref. \citep{rkzg-sdxn}), the multiphoton
input and detection cases such as ($1$, $2$) and ($4$, $1$) configurations
especially $n=2$ case can also obtain very well approximated SC state,
and it will confirm in following related contents. 

To deeply illustrate the effects of OPA on SV state with different
input and output photon configuration characterized by $m$ and $n$,
we investigate the complexity and negativity of our heralded states.

The Wigner negativity of a quantum state $\rho$ is defined as \citep{Kenfack2004}
\begin{equation}
\mathcal{N}=\int_{\mathbb{C}}\vert W(\alpha\vert\rho)\vert\frac{d^{2}\alpha}{\pi}-1,\label{eq:18}
\end{equation}
 where $W(\alpha\vert\rho)$ is the Wigner function of the state $\rho$.
This quantity characterizes the volume of the negative part of the
Wigner function in phase space. It serves as a strong indicator of
non\nobreakdash-classicality: a non\nobreakdash-zero value of $\mathcal{N}$
signals genuine quantum interference, which is a resource for e.g.
CV quantum computing. However, Wigner negativity does not very accurately
capture structural features such as squeezing or high photon number
when the Wigner function remains non\nobreakdash-negative (e.g.,
for SV states).

In contrast, the complexity $\mathcal{C}(\rho)$ introduced by Tang
et al. \citep{Tang_2025} is defined via the always\nobreakdash-positive
Husimi function $Q(\alpha\vert\rho)=\langle\alpha\vert\rho\vert\alpha\rangle$:
\begin{equation}
\mathcal{C}(\rho)=e^{S_{W}(\rho)-1}I(\rho)\label{eq:19-1}
\end{equation}
 with the Wehrl entropy 
\begin{equation}
S_{W}(\rho)=-\int Q(\alpha\vert\rho)\ln Q(\alpha\vert\rho)\frac{d\alpha^{2}}{\pi}
\end{equation}
and the Fisher information 
\begin{equation}
I(\rho)=\frac{1}{4}\int\frac{\vert\vert\nabla Q(\alpha\vert\rho)\vert\vert^{2}}{Q(\alpha\vert\rho)}\frac{d\alpha^{2}}{\pi}.
\end{equation}
The quantity $\mathcal{C}(\rho)$ captures the configurational trade\nobreakdash-off
between the spread (disorder) and localization (order) of the state
in phase space. It is minimal ( $\mathcal{C}=1$) for all displaced
thermal states (including vacuum and coherent states) and increases
with squeezing and photon number. Importantly, complexity can be large
even when the Wigner function is completely non\nobreakdash-negative
(e.g., SV), thus revealing a form of structural richness that is independent
of negativity.

Together, Wigner negativity and complexity provide complementary insights:
the former signals non\nobreakdash-classical interference, whereas
the latter quantifies the phase\nobreakdash-space shape complexity.
Their combined use allows us to distinguish, for example, a SV (high
complexity, zero negativity) from a SC state (high negativity, moderate
complexity) after herald\nobreakdash-based operations.

This complementary perspective becomes particularly valuable when
assessing quantum resources under realistic loss. As we will show
in Sec. \ref{sec:6} (see Fig. \ref{fig:13}), under photon loss,
$\mathcal{C}$ decays much more slowly than Wigner negativity $\mathcal{N}$.
Even when $\mathcal{N}$ vanishes at sufficiently high loss rates,
$\mathcal{C}$ remains significantly above its minimal value of $1$,
indicating that the state retains nontrivial phase-space structure.
Such structural richness can translate into practical advantages:
high $\mathcal{C}$ correlates with large QFI in phase estimation
(see Sec. \ref{sec:5}), and the persistence of $\mathcal{C}$ under
loss suggests potential robustness for encoding information in fault-tolerant
bosonic quantum error correction. Therefore, $\mathcal{C}$ not only
complements $\mathcal{N}$ as a witness of non-classicality, but also
serves as a more loss-resilient indicator of structural complexity
relevant to quantum metrology and error correction.

Fig. \ref{fig:4} shows the variation of the Wigner negativity volume
$\mathcal{N}$ and complexity $\mathcal{C}$ of the heralded quantum
states generated by some different configurations of $m$ and $n$
with respect to the gain parameter $g$, under fixed initial squeezing
parameter $r=1.0$. We can see as follows: (i) As $g$ increases,
the magnitude order of both negative volume $\mathcal{N}$ and complexity
$\mathcal{C}$ gradually increase with increasing of $n$ ($n=4>n=3>n=2$).
This indicates that within this parameter interval, higher-order non-Gaussian
operations such as $\ensuremath{n=3,4}$ can produce the states which
possess stronger nonclassicality and more structural richness. (ii)
For all $\ensuremath{n}$ in the region where $g$ is relatively small,
the negative volume $\mathcal{N}$ increases rapidly with the growth
of $g$. When $g$ increases to a certain extent, the growth of all
curves slows down, eventually stabilizes, and approaches their respective
saturation values, and this trend also hold for complexity $\mathcal{C}$.

\section{\label{sec:4}Examples with specific configurations}

In above we have introduced the general information about multiphoton
detection for SV input case. As concrete examples, in following subsections
we check the concrete details of state $\vert\Psi\rangle_{sv,1,n}$
for $n=2,3,4$ cases and investigate their related features.

\subsection{\label{subsec:1}Two-photon heralding ($n=2$)}

For two-photon detection, $\Pi_{2}=\vert2\rangle\langle2\vert$, we
can obtain the corresponding signal output state as (see Eq. (\ref{eq:9-1})): 

\begin{equation}
\vert\Psi\rangle_{sv,1,2}=\mathcal{N^{\prime}}S(\xi^{\prime\prime})\left[c_{1}\vert1\rangle+c_{3}\vert3\rangle\right]\label{eq:13}
\end{equation}
where $\mathcal{N^{\prime}}=1/\sqrt{\vert c_{1}\vert^{2}+\vert c_{3}\vert^{2}}$
is the normalization constant and

\begin{subequations}
\begin{align}
c_{1} & =-\frac{\sqrt{2}}{g^{2}}G\lambda e^{\frac{i3}{2}\varPhi(z,z')}\sqrt{\cosh r'}A\nonumber \\
 & +\frac{3\sqrt{2}}{2g}G^{3}e^{i\theta}\sinh r\lambda e^{\frac{\phi_{z}(\theta')}{2}+\frac{i3}{2}\varPhi(z,z^{\prime})}(\cosh r')^{\frac{3}{2}}AB\\
c_{3} & =-\frac{\sqrt{3}}{g}G^{3}e^{i\theta}\sinh r\lambda e^{\frac{\phi_{z}(\theta')}{2}+\frac{i7}{2}\varPhi(z,z^{\prime})}(\cosh r')^{\frac{3}{2}}A^{2}\label{eq:34}
\end{align}
\begin{equation}
\zeta_{1}=\tanh\frac{r}{2}e^{i\theta};\zeta_{2}=\tanh\frac{r^{\prime}}{2}e^{i\varphi}\label{eq:19}
\end{equation}

\begin{align}
\varPhi(z,z') & =\frac{1}{i}\ln\left(\frac{1+\zeta_{1}\zeta^{*}_{2}}{1+\zeta^{*}_{1}\zeta_{2}}\right),\ z^{\prime}=r^{\prime}e^{i\varphi}\\
\lambda & =e^{[\ln g+\phi_{z}(\theta^{\prime})]/2}
\end{align}

\begin{equation}
\tanh\frac{r^{\prime\prime}}{2}e^{i\varphi^{\prime\prime}}=\frac{\zeta_{1}+\zeta_{2}}{1+\zeta_{1}\zeta^{*}_{2}};\xi^{\prime\prime}=r^{\prime\prime}e^{i\varphi^{\prime\prime}}.\label{eq:21-1}
\end{equation}
 \end{subequations} 

Here $A$ and $B$ are given as 

\begin{align}
A & =\cosh r\cosh r^{\prime}+e^{-i(\theta-\varphi)}\sinh r\sinh r^{\prime},\nonumber \\
B & =\cosh r\sinh r^{\prime}e^{-i\varphi}+\sinh r\cosh r^{\prime}e^{-i\theta}
\end{align}

It can be seen that the output state is the superposition of squeezed
one- and three-photon number states. For $r\in(0,1]$, the $\vert1\rangle$
component always larger than $\vert3\rangle$ component. The Wigner
function of the output state $\vert\Psi\rangle_{sv,1,2}$ for $r=1.0$
and $g=$1.5 can be seen in the second row of the Fig. \ref{fig:2}.
As investigated in most recent work \citep{hr5f-lvy7}, our heralded
state $\vert\Psi\rangle_{sv,1,2}$ can be used as logical codewords
to correct single-photon loss and dephasing error channels in combination. 

In our work we mainly discuss the similarity of our output state $\vert\Psi\rangle_{sv,1,n}$
to the squeezed odd or even Schr\"odingerr cat ( SOSC, SESC ) states
which defined as: 
\begin{equation}
\vert\Psi_{\theta}\rangle=\mathcal{N}_{\theta}S(\gamma)\vert Cat\rangle_{\theta,\alpha},\label{eq:23}
\end{equation}
 where the normalization coefficient is given by $\mathcal{N}_{\theta}=(2+2e^{-2\alpha^{2}}e^{2r}\cos\theta)^{-1/2}$,
and squeezing parameter $\gamma$ to be real. As previous works shown
\citep{ourjoumtsev2007generation,PhysRevA.78.063811}, $k$-photon
subtraction from the SV state, i.e., $a^{k}S(z)\vert0\rangle$, can
provide a well approximation to a certain type of squeezed SC state
$S(\gamma)\vert Cat\rangle_{k\pi,\alpha}$. In those protocols, achievable
amplitude is at most $\alpha\sim\sqrt{k}$ while maintaining high
fidelity. For the experimental realization of $k$-photon subtraction
(up to $4$-photons to date ), please see the references listed in
Table \ref{tab:9-1}. Since the squeezing operation, when properly
adjusted, reduces the average photon number of the SC state, the squeezed
SC states can survive longer than their conventional counterparts
in a lossy environment. This property makes squeezed SC states promising
for a wide range of applications in quantum computing, including quantum
error correction \citep{PhysRevX.10.011058,PhysRevA.106.022431}. 

In order to confirm the assumption, the fidelity between our output
state $\vert\Psi\rangle_{sv,1,2}$ and target SOSC state $\vert\Psi_{\pi}\rangle$
can be checked as 
\begin{align}
F_{1} & =\vert\langle\Psi_{\pi}\vert\Psi\rangle_{sv,1,2}\vert^{2}\nonumber \\
 & =\frac{16}{9}\left(\nu\chi\right)^{2}\exp\left(\frac{-2\alpha^{2}}{(1+g^{2})}\right)\times\nonumber \\
 & \vert\frac{g^{\frac{3}{2}}\alpha(3\sqrt{2}(1+g^{2})^{2}+\sqrt{3}(3-3g^{4}+4g^{2}\alpha^{2})\kappa^{*})}{(1+g^{2})^{\frac{7}{2}}}\vert^{2}\label{eq:21}
\end{align}
with $\nu=(2-2e^{-2\alpha^{2}})^{-\frac{1}{2}}$ , $\chi=(1+\vert\kappa\vert^{2})^{-\frac{1}{2}}$
and $\kappa=c_{3}/c_{1}$. 

\begin{figure}
\includegraphics[width=8cm]{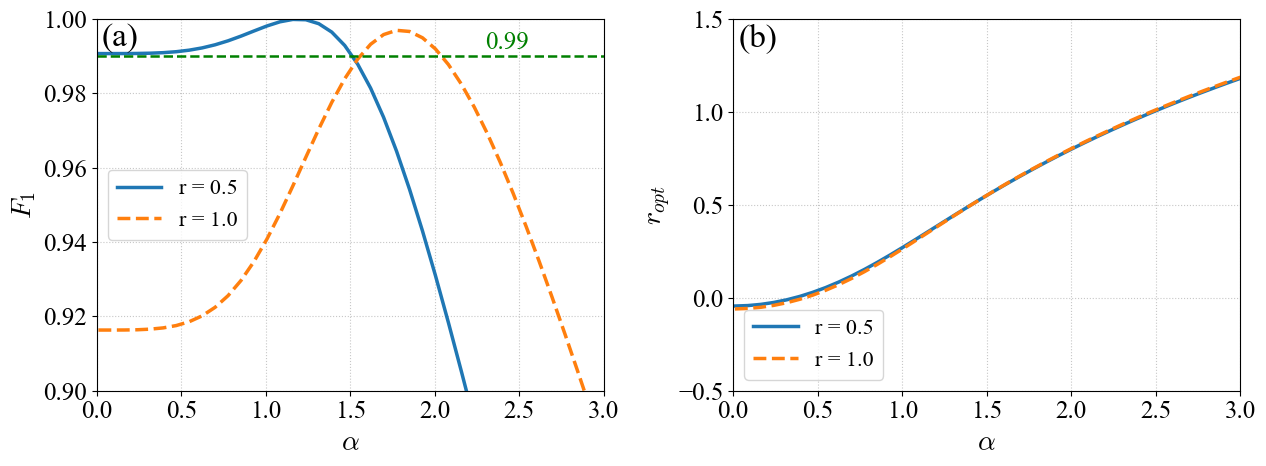}

\caption{\label{fig:3}(a) Optimal fidelity between the output state $\vert\Psi\rangle_{sv,1,2}$
and the target SOSC state $\vert\Psi_{\pi}\rangle$ as a function
of coherent amplitude of the target state for different squeezing
parameters $r$. (b) Corresponding squeezing parameter $\gamma$ of
the target SOSC state that maximizes the fidelity in (a). The other
parameters are the same as in Fig. \ref{fig:2}.}
\end{figure}

In Fig. \ref{fig:3}(a), the optimized fidelity $F_{1}$ corresponding
the Eq. (\ref{eq:21}) as a function of coherent amplitude $\alpha$
of target state is plotted for different squeezing parameters $r=0.5$
($\approx4.34\ dB$), $r=1.0$ ($\approx8.69\ dB$) and the parameter
$g$ is set to $1.5$. In Fig. \ref{fig:3} (b) the corresponding
squeezing parameter $\gamma$ of the target SOSC state that can give
the optimized fidelity values in Fig. \ref{fig:3} (a) is showed.
As can be observed, when $r=0.5$ is considered, the nearly perfect
fidelity $F_{max}=0.999$ can be achieved for $\alpha=1.191$ and
$\gamma_{opt}=0.378$. Furthermore, in $r=1.0$ case, the maximum
fidelity $F_{max}=0.997$ is occurred to the corresponded $\alpha=1.795$
and $\gamma_{opt}=0.706$ {[} see Fig. \ref{fig:3}(b) {]}. It's also
interesting to note that for $r=1.0$ case, the range of amplitude
$\alpha$ for $F_{1}>0.99$ is from $1.565$ to $2.029$. 
\begin{table}
\centering
\caption{\label{tab:1}Optimized parameters of the fidelity between the output
state $\vert\Psi^{\prime}\rangle_{sv,1,2}$ and target state $\vert\Psi_{\pi}\rangle$
when the squeezing parameter of the input SV is fixed to $r=0.74$.
$\alpha=1.732$}

\begin{tabular*}{8cm}{@{\extracolsep{\fill}}ccc}
\toprule 
$g$ & $\gamma$ & $F_{1}$\tabularnewline
\multicolumn{1}{c}{1.05} & 1.276 & 0.934\tabularnewline
1.1 & 1.133 & 0.945\tabularnewline
1.5 & 0.668 & 0.993\tabularnewline
2.0 & 0.505 & 0.998\tabularnewline
2.5 & 0.437 & 0.998\tabularnewline
5 & 0.350 & 0.997\tabularnewline
10 & 0.329 & 0.997\tabularnewline
\bottomrule
\end{tabular*}
\end{table}

In Table \ref{tab:1}, we give the optimized parameters and corresponding
fidelity $F_{1}$ for fixed squeezing parameter $r=0.74$ and fixed
amplitude $\alpha=1.732$ while changing the gain values $g$ from
$1.01$ to $10$. As $g$ increases, the optimal value of $\gamma$
decreases, As we noticed, the optimal fidelity $F_{1}$ for the entire
range of $g$ remains considerable high values, $F_{1}\ge0.993$,
and reached the saturating value $0.997$ for $g\ge5.0$. The numerical
data in Table \ref{tab:1} indicated that our protocol with ($1$,
$2$) configuration can be equivalent to $3$-photon subtraction from
SV state, since $\sqrt{3}\approx1.732$. 

\begin{figure}
\includegraphics[width=8cm]{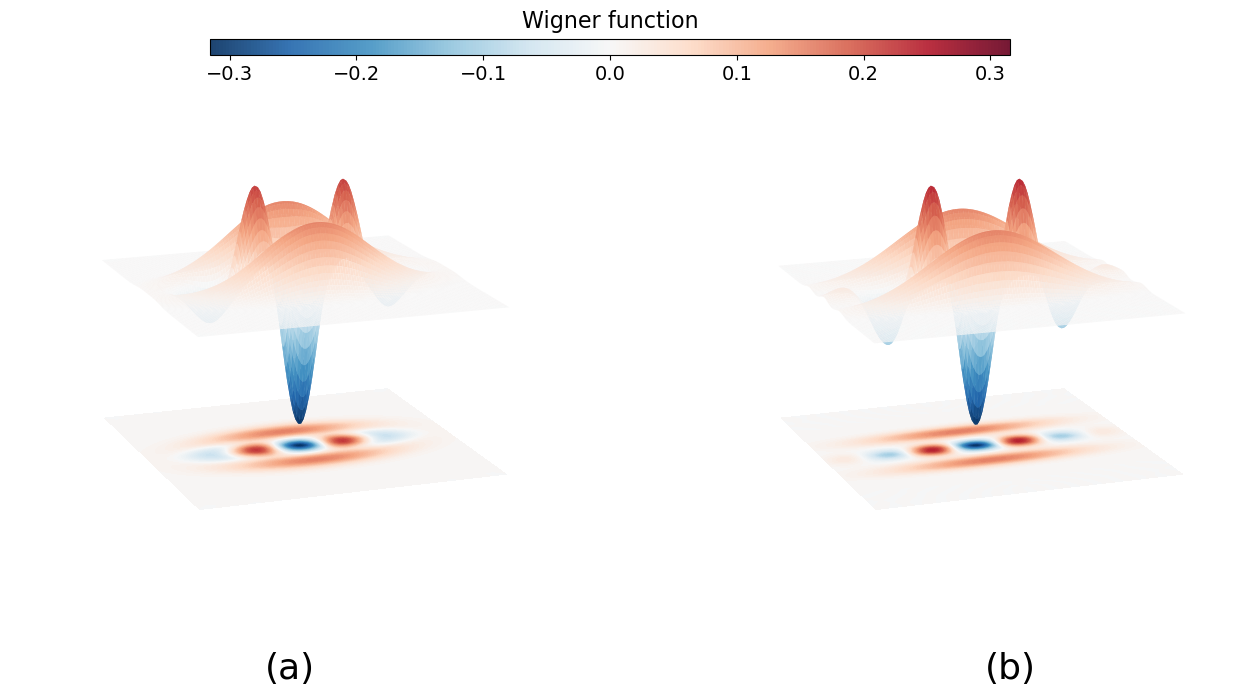}

\caption{\label{fig:6}(a) Wigner function of the generated state with $r=1.0$
and $g=1.5$. (b) Wigner function of the ideal SOCS state with $\gamma=0.803$
and $\alpha=2.0$. Other parameters are the same as those used in
Fig. \ref{fig:3}. For these parameters, the fidelity calculated using
the Wigner functions of the generated and ideal SOCS state is $F_{1}=0.992$.The
other parameters are the same as in Fig. \ref{fig:2}.}
\end{figure}

Figure. \ref{fig:6} (a) displays the Wigner function of the heralded
state $\vert\Psi\rangle_{sv,1,2}$ for $r=1.0$ and $g=1.5$. The
ideal target SOSC state with $\gamma=0.803$ and $\alpha=2.0$ is
shown in Figs. \ref{fig:3} (b) yielding a high fidelity $F_{1}=0.992$
to the our heralded state $\vert\Psi\rangle_{sv,1,2}$. The high similarity
is evident from the characteristic negative Wigner region and the
two positive peaks. 

To explore the limits of our protocol, we further optimized the fidelity
for a target SOSC state with an even larger amplitude $\alpha=2.236$
(corresponding to $\sqrt{5}$) for regions $r\in[1,3]$ and $g\in(1,10]$
and found that the fidelity always in $0.983\pm0.002$ level. While
this is slightly lower than the optimal fidelity achieved at $\alpha=1.732$
(see Table \ref{tab:1}, where $F_{1}\ge0.993$), it still represents
a highly competitive level of quantum state fidelity for SC states
in this amplitude regime. Achieving $\alpha\ge2$ is critical for
quantum information, as it ensures near-orthogonality between the
two cat-state components ($e^{-2\alpha^{2}}\ll1$), enabling fault-tolerant
bosonic quantum error correction \citep{hr5f-lvy7}.

Compared to traditional multiphoton subtraction, which requires detecting
at least four photons to reach $\alpha\sim2$, our scheme achieves
the same with only two-photon detection and a corresponding single-trial
successful probability on the order of $10^{-2}$ (see Sec. \ref{sec:6}
for details), significantly enhancing the success probability (scaling
as $R^{2}$ vs $R^{4}$, where $R\sim0.02-0.05$ is the beam-splitter's
reflectivity in conventional subtraction schemes). Moreover, our single-OPA
setup is more compact than GPS schemes requiring two squeezed sources
and a beam splitter. These results confirm that our two-photon heralding
protocol reliably generates high-fidelity SOSC states with amplitudes
$\alpha\gtrsim2$, reaching a regime essential for fault-tolerant
quantum computing while offering experimental simplicity and higher
generation rates.

\begin{figure}
\centering
\includegraphics[width=8cm]{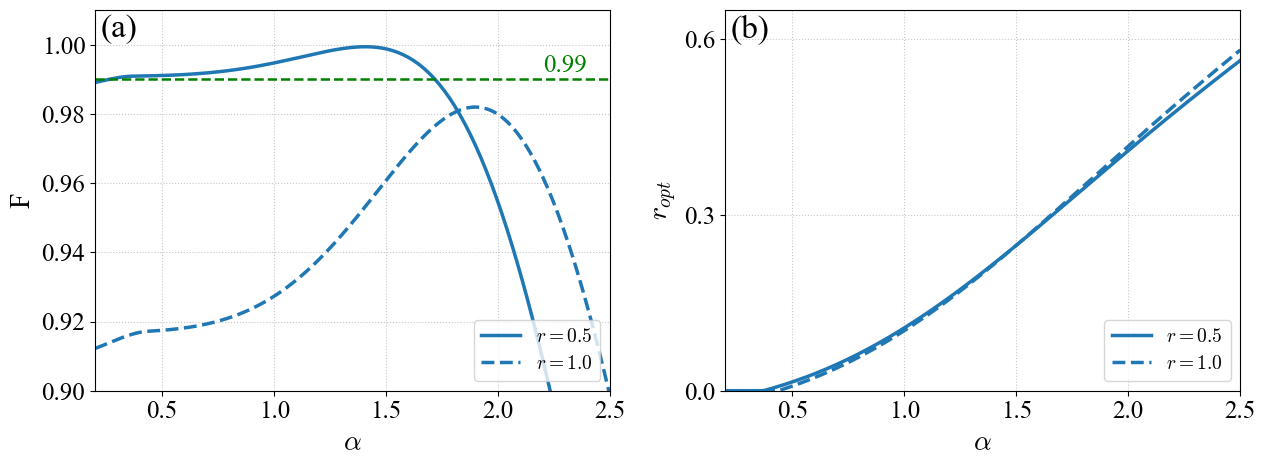}

\caption{\label{fig:7} (a) Optimized parameters for the fidelity between $\vert\Psi\rangle_{sv,1,2}$
with $a^{2}\vert\Psi_{\pi}\rangle$ for $r=0.5$ (solid curve) and
$r=1.0$ (dashed curve). (b) Corresponding squeezing parameter $\gamma$
of the target state that maximizes the fidelity in (a). The other
parameters are the same as in Fig. \ref{fig:2}. }
\end{figure}

In addition to the ideal SOSC state, the heralded state $\vert\Psi\rangle_{sv,1,2}$
generated by two-photon detection exhibits excellent agreement with
the target states $a^{2}\vert\Psi_{\pi}\rangle.$ Figure. \ref{fig:7}
(a) presents the optimized fidelity as a function of the coherent
amplitude $\alpha$ of the target states, while Fig. \ref{fig:7}
(b) shows the corresponding optimal squeezing parameter $\gamma$
of the target state. For the parameters $g=1.5$ and $r=0.5$ ($1.0$),
the optimized fidelity between $\vert\Psi\rangle_{sv,1,2}$ and $a^{2}\vert\Psi_{\pi}\rangle$
reaches $F=0.999$ ($0.982$) at $\alpha=1.399$ ($1.900$). For the
$r=0.5$, we observe a broad high-fidelity region ( $F>0.99$) for
$\alpha$ ranging from $0.01$ to $1.701$. In contrast, when $r=1.0$,
the optimal fidelity always lower then $0.99$ for entire parameter
regimes. In both scenarios, the optimized coherent amplitude $\alpha$
increases with the gain parameter $g$. 

\subsection{\label{subsec:2}Three-photon heralding ($n=3$) }

We now consider the case where the idler output is projected onto
the Fock state $\vert3\rangle\langle3\vert$, i.e., $\hat{\Pi}=|3\rangle_{b}\langle3\vert$.
Substituting $n=3$ into the general expression Eq. (\ref{eq:9-1})
for $m=1$ cases, we obtain the following normalized output state
for the signal mode: 

\begin{equation}
\vert\Psi\rangle_{sv,1,3}=\mathcal{N^{\prime\prime}}S(\xi^{\prime\prime})\left[c_{0}\vert0\rangle+c_{2}\vert2\rangle+c_{4}\vert4\rangle\right]\label{eq:26}
\end{equation}
where $\mathcal{N^{\prime\prime}}=1/\sqrt{\vert c_{0}\vert^{2}+\vert c_{2}\vert^{2}+\vert c_{4}\vert^{2}}$
is the normalization constant. The complex squeezing parameter $\xi^{\prime\prime}$
in this final squeezing operator $S(\xi^{\prime\prime})$ is defined
in Eq. (\ref{eq:21-1}), and the coefficients $c_{0}$, $c_{2}$,
$c_{4}$ are given by \begin{subequations}
\begin{align}
c_{0} & =-\frac{\sqrt{6}}{2g^{2}}G^{2}\lambda e^{\frac{i}{2}\varPhi(z,z')}\sqrt{\cosh r'}AB+\nonumber \\
 & \frac{\sqrt{6}}{2g}G^{4}e^{i\theta}\sinh r\lambda e^{\frac{\phi_{z}(\theta')}{2}+\frac{i}{2}\varPhi(z,z')}(\cosh r')^{\frac{3}{2}}\sinh rAB^{2},\\
c_{2} & =\frac{\sqrt{3}}{g^{2}}G^{2}\lambda e^{\frac{i5}{2}\varPhi(z,z')}\sqrt{\cosh r'}A^{2}-\nonumber \\
 & \frac{2\sqrt{3}}{g}G^{4}e^{i\theta}\sinh r\lambda e^{\frac{\phi_{z}(\theta')}{2}+\frac{i5}{2}\varPhi(z,z')}(\cosh r')^{\frac{3}{2}}\sinh rA^{2}B,\\
c_{4} & =\frac{2}{g}G^{4}e^{i\theta}\sinh r\lambda e^{\frac{\phi_{z}(\theta')}{2}+\frac{i9}{2}\varPhi(z,z')}(\cosh r')^{\frac{3}{2}}\sinh rA^{3}\label{eq:39}
\end{align}
 \end{subequations} The symbols $A$, $B$, $\varPhi(z,z')$, $\phi_{z}(\theta')$
and $\lambda$ are defined in Eqs. (\ref{eq:19}-\ref{eq:21-1}). 

From Eq. (\ref{eq:26}) we observe that the heralded state contains
only even Fock components $\vert0\rangle$, $\vert2\rangle$ and $\vert4\rangle$.
This is consistent with the parity argument given in above content:
for odd $n$ ( here $n=3$ ), the output state retains the even-parity
structure of the input SV. The $\vert0\rangle$ component is generally
very small compared to $\vert2\rangle$ and $\vert4\rangle$ for the
parameter ranges considered in this work. The Wigner fucntion of $\vert\Psi\rangle_{sv,1,3}$
for a representative set of parameters ($r=1.0$ and $g=1.5$ ) is
shown in this second row and forth column of Fig. \ref{fig:2}. Since
the state contains only even Fock components, the Wigner function
exhibits a positive central peak surrounded by negative regions at
larger phase-space radii. This structure is characteristic of even-parity
non-Gaussian states and differs markedly from the odd cat-like states
obtained for $n=2$. 

Over the entire parameter ranges we explored ( $r\in\left[0,3\right]$
and $g\in(1,10]$), the fidelity of the ($1$, $3$) configuration
with an ideal SESC state never exceeds $0.914$, which is substantially
below the $0.970\pm0.010$ level achieved by ($3$,$1$) configuration
(see Table \ref{tab:9-1}). Thus ($1,$ $3$) is not a viable choice
for preparing high-fidelity SESC state with amplitude $\alpha\sim2$
(equivalent to four-photon subtraction). Instead, the ($1$, $3$)
configuration should be preferred for such tasks.

\begin{figure}
\includegraphics[width=8cm]{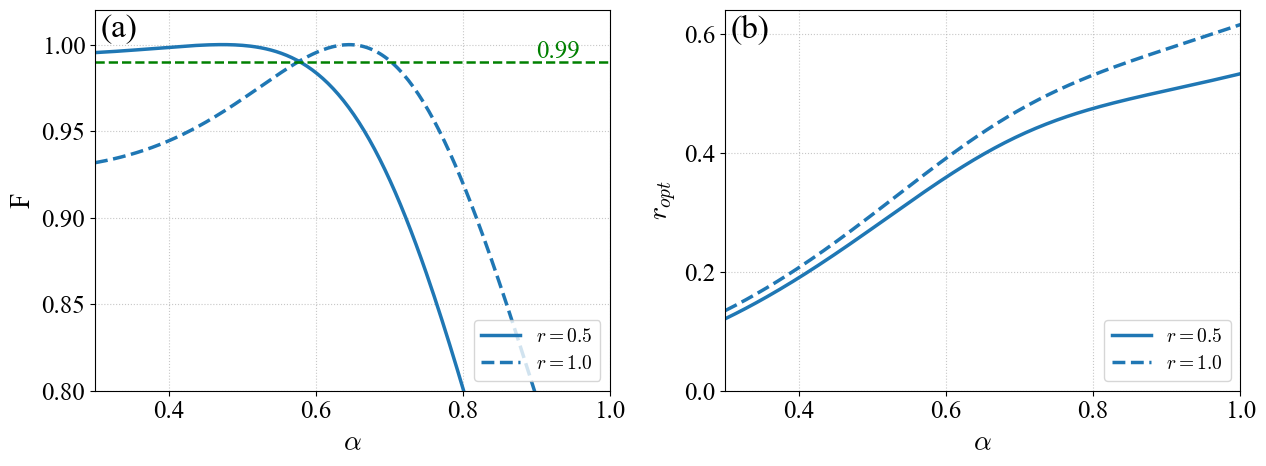}

\caption{\label{fig:8}(a) Optimized parameters for the fidelity between $\vert\Psi\rangle_{sv,1,3}$
with $a^{\dagger2}\vert\Psi_{0}\rangle$ for $r=0.5$ (solid curve)
and $r=1.0$ (dashed curve). (b) Corresponding squeezing parameter
$r$ of the target state that maximizes the fidelity in (a). The other
parameters are the same as in Fig. \ref{fig:2}.}
\end{figure}

After careful consideration, we find that $\vert\Psi\rangle_{sv,1,3}$
exhibits high fidelity with the target state $a^{\dagger2}\vert\Psi_{0}\rangle$.
To quantify the non\nobreakdash-Gaussian character of the three\nobreakdash-photon
heralded state $\vert\Psi\rangle_{sv,1,3}$ given in Eq. (\ref{eq:26}),
we evaluate its overlap with the even\nobreakdash-parity target state
$a^{\dagger2}\vert\Psi_{0}\rangle$ that naturally arise from photon
operations on squeezed SC state. Figure \ref{fig:8} shows the optimized
fidelity as a function of the target coherent amplitude $\alpha$
for two input squeezing parameters, $r=0.5$ and $r=1.0$, at fixed
OPA gain $g=1.5$. For $r=0.5$, the fidelity can exceed $0.99$ over
a broad range of $\alpha$ and the maximum fidelity reach nearly $0.999$.
When $r$ increases to $1.0$ the optimal Fidelity shifts to larger
$\alpha$, while the peak fidelity remains above $0.99$. Panel (b)
displays the corresponding optimal squeezing parameter $\gamma$ of
the target states that maximizes the fidelity in panel (a). The $\gamma_{opt}$
increases monotonically with increasing $\alpha$ and $r=0.5$ case
generally smaller for the larger input squeezing $r=1.0$. 

\begin{table}
\centering
\caption{\label{tab:1-4-1}Optimized parameters for the fidelity between $\vert\Psi\rangle_{sv,1,3}$
with $|\phi_{1}\rangle$}

\begin{tabular*}{8cm}{@{\extracolsep{\fill}}ccccc}
\toprule 
$r$ & $g$ & $\gamma_{opt}$ & $\alpha$ & $F$\tabularnewline
\midrule
\multirow{4}{*}{0.5} & 1.5 & 0.250 & 0.473 & \multirow{8}{*}{0.999}\tabularnewline
 & 2.5 & 0.133 & 0.586 & \tabularnewline
 & 5 & 0.085 & 0.604 & \tabularnewline
 & 10 & 0.073 & 0.612 & \tabularnewline
\cmidrule{1-4}
\multirow{4}{*}{1.0} & 1.5 & 0.430 & 0.645 & \tabularnewline
 & 2.5 & 0.233 & 0.742 & \tabularnewline
 & 5 & 0.141 & 0.780 & \tabularnewline
 & 10 & 0.120 & 0.789 & \tabularnewline
\bottomrule
\end{tabular*}
\end{table}

In order to further confirm the similarities of the between our heralded
state $\vert\Psi\rangle_{sv,1,3}$ and $\vert\phi_{1}\rangle=a^{\dagger2}\vert\Psi_{0}\rangle$,
we list the optimized fidelities and parameters in Table \ref{tab:1-4-1}.
For input squeezing $r=0.5$ the fidelity reaches $0.999$ for all
$g$ values from $1.5$ to $10$; the optimal coherent amplitude $\alpha$
lies between $0.473$ and $0.612$ , while $\gamma_{opt}$ decreases
from $0.250$ to $0.073$. For $r=1.0$ , the fidelity remains above
$0.998$ with $\alpha\approx0.645-0.789$ and $\gamma_{opt}$ ranging
from $0.430$ to $0.120$. These results demonstrate an excellent
agreement with the two\nobreakdash-photon\nobreakdash-added SESC
state. 

We also notice that the $\vert\phi_{2}\rangle=a^{\dagger}a\vert\Psi_{0}\rangle$
can also give well approximation to our state with ($1$, 3) configuration.
The extreme high fidelity $F\ge0.999$ between $\vert\Psi\rangle_{sv,1,3}$
and the target $\vert\phi_{2}\rangle$ is achieved for all considered
OPA gain value $g\in(1,10]$ with optimal $\alpha\in[1.063,1.324]$
and $\gamma_{opt}\in\left[0.219,0.158\right]$ when the input squeezing
is $r=0.5$. For $r=1.0$, the fidelity is slightly lower ( $0.990-0.996$),
but the achieved coherent amplitudes are significantly larger ( $\alpha\approx1.48-1.69$).
Thus, $\vert\phi_{1}\rangle$ provides higher fidelity, while $\vert\phi_{2}\rangle$
yields a larger SC size, offering a flexible choice depending on the
experimental goal.

The high fidelity between $\vert\Psi\rangle_{sv,1,3}$ and the states
$\vert\phi_{1}\rangle$ and $\vert\phi_{2}\rangle$ can be explained
as follows. First, both target states contain only even Fock components,
matching the parity of our heralded state (for odd $n=3$, the output
state retains the even parity of the input SV). Second, although the
output state has a small $\vert0\rangle$ component, its amplitude
is negligible compared to those of $\vert2\rangle$ and $\vert4\rangle$.
Third, as shown in Table \ref{tab:1-4-1}, the optimal squeezing parameters
$\gamma_{opt}$ are small and decrease with increasing the OPA gain
$g$ (this trend is same to the $\vert\phi_{2}\rangle$ case). For
such small $\gamma$, higher even Fock states ( $\vert6\rangle$,
$\vert8\rangle$,...) in the SCS state are substantially suppressed.
Consequently, both the target states and the heralded state are dominated
by the same low-lying even Fock components ( $\vert2\rangle$ and
$\vert4\rangle$), which explains the observed high fidelities.

\subsection{\label{subsec:3}Four-photon heralding ($n=4$) }

We now turn to the case of detecting four photons at the idler output,
i.e., $\hat{\Pi}=|4\rangle\langle4\vert$). Inserting $n=4$ into
the general formula Eq. (\ref{eq:9-2}) ( with $m=1$ fixed) yields
the heralded signal state

\begin{equation}
\vert\Psi\rangle_{sv,1,4}=\mathcal{N^{\prime\prime\prime}}S(\xi^{\prime\prime})\left[c_{1}\vert1\rangle+c_{3}\vert3\rangle+c_{5}\vert5\rangle\right]
\end{equation}
where $\mathcal{N^{\prime\prime\prime}}=1/\sqrt{\vert c_{1}\vert^{2}+\vert c_{3}\vert^{2}+\vert c_{5}\vert^{2}}$
denotes the normalization constant, and \begin{subequations}
\begin{align}
c_{1} & =\frac{\sqrt{6}}{g^{2}}G^{3}\lambda e^{\frac{3i}{2}\varPhi(\xi,\xi')}\sqrt{\cosh r'}A^{2}B\nonumber \\
 & +\frac{15\sqrt{6}}{12}G^{5}e^{i\theta}\sinh r\lambda e^{\frac{\phi_{z}(\theta')}{2}+\frac{3i}{2}\varPhi(\xi,\xi')}(\cosh r')^{\frac{3}{2}}A^{2}B^{2}\\
c_{3} & =-\frac{2}{g^{2}}G^{3}\lambda e^{\frac{7i}{2}\varPhi(\xi,\xi')}\sqrt{\cosh r'}A^{3}\nonumber \\
 & -5A^{3}BG^{5}e^{i\theta}\sinh r\lambda e^{\frac{\phi_{z}(\theta')}{2}+\frac{7i}{2}\varPhi(\xi,\xi')}(\cosh r')^{\frac{3}{2}}\\
c_{5} & =\sqrt{5}G^{5}e^{i\theta}\sinh r\lambda e^{\frac{\phi_{z}(\theta')}{2}+i\frac{11}{2}\varPhi(\xi,\xi')}(\cosh r')^{\frac{3}{2}}A^{4}.\label{eq:45}
\end{align}
 \end{subequations}A striking feature is that only odd Fock states
appear --- a direct consequence of parity conservation: for even
$n$ (here $n=4$), the output state acquires odd parity relative
to the input SV. The corresponding Wigner function, plotted in the
fourth row fourth comlumn of Fig. \ref{fig:2} for $r=1.0$ and $g=1.5$,
shows a deep negative dip at the origin and two well\nobreakdash-separated
positive humps, reminiscent of an odd SC state but with richer structure
due to the $\vert5\rangle$ contribution. We notice that our heralded
state $\vert\Psi\rangle_{sv,1,4}$ also cannot be approximated to
a SOSC state, as its maximum fidelity is only $0.816$ over all parameter
regimes we considered ( $r\in\left[0,3\right]$ and $g\in(1,10]$). 

By numerical calculation we have found that there are lots of states
such as $\vert\psi_{1}\rangle=a^{\dagger2}\vert\Psi_{\pi}\rangle$,
$\vert\psi_{2}\rangle=a^{\dagger3}a\vert\Psi_{\pi}\rangle$ and $\vert\psi_{3}\rangle=a^{\dagger2}a^{2}\vert\Psi_{\pi}\rangle$
can give extreme similarities with our heralded state $\vert\Psi\rangle_{sv,1,4}$.
Each of these contains only odd Fock components, matching the parity
of our heralded state. As an concrete example, below we choose $\vert\psi_{1}\rangle$
to give the analysis about the optimal fidelity between $\vert\Psi\rangle_{sv,1,4}$
and $\vert\psi_{1}\rangle$. 

\begin{figure}
\includegraphics[width=8cm]{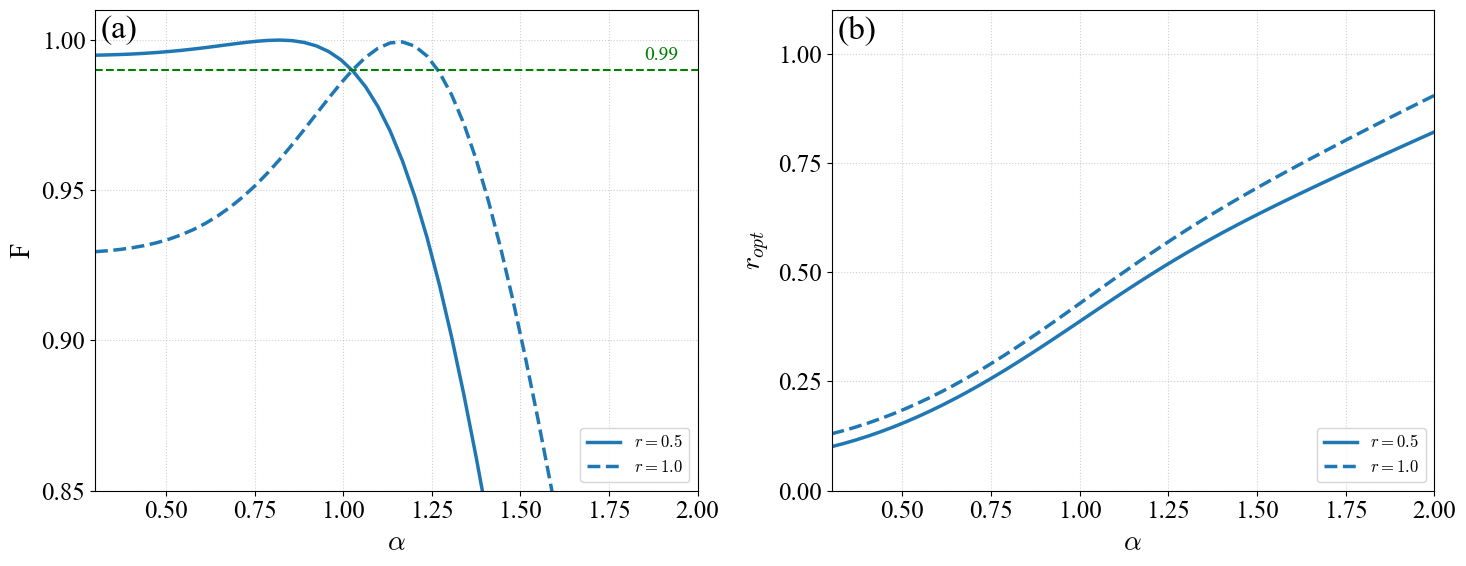}

\caption{\label{fig:9}(a) Optimized parameters for the fidelity between $\vert\Psi\rangle_{sv,1,4}$
with $a^{\dagger2}\vert\Psi_{\pi}\rangle$ for $r=0.5$ (solid curve)
and $r=1.0$ (dashed curve). (b) Corresponding squeezing parameter
$r$ of the target state that maximizes the fidelity in (a). The other
parameters are the same as in Fig. \ref{fig:2}.}
\end{figure}

Figure\,\ref{fig:9} (a) illustrates the optimized fidelity between
the heralded state $\vert\Psi\rangle_{sv,1,4}$ and the target state
$\vert\psi_{1}\rangle$ as a function of the target coherent amplitude
$\alpha$ for input squeezing parameters $r=0.5$ (solid curves) and
$r=1.0$ (dashed curves), with OPA gain fixed at $g=1.5$. For $r=0.5$,
all the fidelity exceed $0.99$ over a certain range of $\alpha$.
When $r$ increases to $1.0$, the optimal $\alpha$ shifts to larger
values, and the optimal fidelity can reach above $0.99$ within a
very narrow region. The corresponding optimal squeezing parameter
$\gamma_{_{opt}}$of the target states versus $\alpha$ is shown in
Fig.\,\ref{fig:9} (b). The similarity between our signal output state
and $\vert\psi_{1}\rangle$ can be further confirmed by the data presented
in Tables \ref{tab:5}. The data are shown for input squeezings $r=1.0$
and $r=1.0$ over a range of OPA gains $g=1.5$ to $10.$ For $r=0.5$,
the fidelity of $0.999$ or higher across the entire gain range, indicating
an excellent match. The target $\vert\psi_{1}\rangle$ yields a moderate
coherent amplitude $\alpha$ (from $0.891$ to $1.049$) with very
high fidelity. Furthermore, for the larger input squeezing $r=1.0$
the fidelity remains highly competitive. The fidelity stays above
$0.998$ with $\alpha$ ranging from $1.157$ to $1.354$. 

\begin{table}
\centering
\caption{\label{tab:5}Optimized parameters for the fidelity between the heralded
state $\vert\Psi\rangle_{sv,1,4}$ and the target state $\psi_{1}\rangle$.The
input squeezing $r$ and OPA gain $g$ are varied; $\gamma_{opt}$
\LyXZeroWidthSpace{} and $\alpha$ are the optimal squeezing and coherent
amplitude of the target state, respectively, and $F$ is the corresponding
fidelity. }

\begin{tabular*}{8cm}{@{\extracolsep{\fill}}ccccc}
\toprule 
$r$ & $g$ & $\gamma_{opt}$ & $\alpha$ & $F$\tabularnewline
\midrule
\multirow{4}{*}{0.5} & 1.5 & 0.291 & 0.891 & \multirow{4}{*}{0.999}\tabularnewline
 & 2.5 & 0.191 & 0.979 & \tabularnewline
 & 5 & 0.148 & 1.035 & \tabularnewline
 & 10 & 0.137 & 1.049 & \tabularnewline
\midrule
\multirow{4}{*}{1.0} & 1.5 & 0.518 & 1.157 & \multirow{4}{*}{0.998}\tabularnewline
 & 2.5 & 0.325 & 1.300 & \tabularnewline
 & 5 & 0.243 & 1.344 & \tabularnewline
 & 10 & 0.222 & 1.354 & \tabularnewline
\bottomrule
\end{tabular*}

\end{table}

For comparison, we also take fidelity analysis between $\vert\Psi\rangle_{sv,1,4}$
and other two target states $\vert\psi_{2}\rangle$ and $\vert\psi_{3}\rangle$.
For $\vert\psi_{2}\rangle$, the fidelity varies between $0.983$
and $0.999$ with $\alpha$ around $0.74-0.79$. Meanwhile, the fidelity
for $\vert\psi_{3}\rangle$ is $0.998\pm0.001$, and it produces the
largest coherent amplitudes among the three target states, reaching
$\alpha\approx1.55$ at $g=10$. The choice of target state thus depends
on the experimental priority. If obtaining a larger coherent amplitude
$\alpha$ is the primary objective, the scheme using $a^{\dagger2}a^{2}S(\gamma)\vert Cat\rangle_{\pi,\alpha}$
(i.e., $\vert\psi_{3}\rangle$) exhibits the best performance; conversely,
if a higher fidelity is required, adopting $a^{\dagger3}aS(\gamma)\vert Cat\rangle_{\pi,\alpha}$
(i.e., $\vert\psi_{2}\rangle$) shows a significant advantage. Furthermore,
$a^{\dagger2}S(\gamma)\vert Cat\rangle_{\pi,\alpha}$ (i.e., $\vert\psi_{1}\rangle$)
provides an excellent compromise that balances both the amplitude
and high fidelity. These trends are clearly visible in the tables:
$\vert\psi_{2}\rangle$ consistently gives the highest fidelity (especially
for $r=0.5$), $\vert\psi_{3}\rangle$ yields the largest $\alpha$,
and $\vert\psi_{1}\rangle$ sits in between with a balanced performance.
Overall, the four-photon heralded state can be approximated with very
high fidelity by any of these three target states, offering flexible
trade-offs for different quantum information tasks.

\subsection{\label{subsec:4}Four-photon input and single-photon detection ($m=4$,
$n=1$) and other cases }

In the previous subsections we fixed the idler input to a single photon
($m=1$) and varied the detected photon number $n=2,3,4$. For $n=2$,
the output state approximates a squeezed odd cat state with amplitude
$\alpha\sim\sqrt{3}\approx1.732$(see Sect. \ref{sec:2} A ) . For
$n=3$ and $n=4$, the heralded states are even- and odd-parity Fock
superpositions ( $\vert0\rangle+\vert2\rangle+\vert4\rangle$ and
$\vert1\rangle+\vert3\rangle+\vert5\rangle$), respectively, and analyzed
in Sect. \ref{sec:2} B and Sect. \ref{sec:2} C. 

A natural complementary question is whether one can instead inject
more photons into the idler while detecting fewer photons at the output,
and whether such a configuration can also generate large-amplitude
SC states. Here we address this question by studying the case $m=4$
(four photons at the idler input) and $n=1$ (single-photon detection
at the idler output). As will be shown, this configuration yields
a SOSC state with coherent amplitude $\alpha\approx2.24$, corresponding
to an effective five-photon subtraction from the SV.

Substituting $m=4$, $n=1$ into the general formula Eq. (\ref{eq:3})
and using the SV input $\vert\phi\rangle=\vert\xi\rangle$, we obtain
the unnormalized heralded state 
\begin{equation}
\vert\Psi\rangle_{sv,4,1}=\sum^{4}_{k=0}H(k,4,1)(a^{\dagger})^{k-3}g^{-a^{\dagger}a}a^{k}\vert\xi\rangle\label{eq:39-2}
\end{equation}
 where the coefficient $H(k,4,1)$ is defined in Eq. (\ref{eq:4}).
In this summation only the terms with $k=3$ and $k=4$ survive because
the factorial ($k-3$)! in the denominator is defined only for non\nobreakdash-negative
arguments. After evaluating the coefficients and using the properties
of the squeezing operator, the state can be rewritten as a superposition
of single\nobreakdash-, three\nobreakdash-, and five\nobreakdash-photon
added attenuated squeezed states:
\begin{equation}
\vert\Psi\rangle_{sv,4,1}=c^{\prime}_{1}a^{\dagger}\vert\frac{\xi}{g^{2}}\rangle+c^{\prime}_{3}a^{\dagger3}\vert\frac{\xi}{g^{2}}\rangle+c^{\prime}_{5}a^{\dagger5}\vert\frac{\xi}{g^{2}}\rangle,\label{eq:40-1}
\end{equation}
 where $\vert\xi/g^{2}\rangle$ denotes a SV state with the scaled
squeezing parameter $\xi/g^{2}$, and the coefficients $c^{\prime}_{1}$,
$c^{\prime}_{3}$ $c^{\prime}_{5}$ are given by \begin{subequations}
\begin{align}
c^{\prime}_{1} & =3\xi^{2}\left(g^{-1}c^{\prime\prime}_{3}+c^{\prime\prime}_{4}\right),\\
c^{\prime}_{3} & =\frac{\xi^{3}}{g^{2}}\left(g^{-1}c^{\prime\prime}_{3}+6c^{\prime\prime}_{4}\right),\\
c^{\prime}_{5} & =\frac{\xi^{4}}{g^{4}}c^{\prime\prime}_{4}
\end{align}
 \end{subequations}with $c^{\prime\prime}_{3}=H(3,4,1)$ and $c^{\prime\prime}_{4}=H(4,4,1)$.
In above derivation, we use 
\begin{align}
a\vert\xi\rangle & =\xi a^{\dagger}\vert\xi\rangle,\\
a^{2}\vert\xi\rangle & =\xi\vert\xi\rangle+\xi^{2}a^{\dagger2}\vert\xi\rangle.
\end{align}
 A more compact representation that reveals the structure of photon
subtraction is 
\begin{align}
\vert\Psi\rangle_{sv,4,1} & =g^{-a^{\dagger}a}\left(c^{\prime\prime}_{3}+c^{\prime\prime}_{4}ga^{\dagger}a\right)a^{3}\vert\xi\rangle.\label{eq:42}
\end{align}

This expression shows that the heralded state is a linear combination
of $3$\nobreakdash-photon and $4$\nobreakdash-photon subtracted
SV, followed by the noiseless attenuation operator $g^{-a^{\dagger}a}$
and an additional photon creation (embedded in the $a^{\dagger}a$
factor in the second term). 

From the above equation one can deduce that the state $\vert\Psi\rangle_{sv,4,1}$
has odd parity and it can be approximated to a SOSC state with large
coherent amplitude $\alpha$. The reason is as follows. As we know,
the $3$-photon subtracted SV state $a^{3}\vert\xi\rangle$ can be
well approximated by a SOSC state with coherent amplitude at most
$\alpha\sim\sqrt{3}=1.732$. However, in Eq. (\ref{eq:42}) the operator
$a^{\dagger}a$ (multiplied by $c^{\prime\prime}_{4}g$) acts on this
state, and the photon number operator $a^{\dagger}a$ plays the role
of amplifying the corresponding SC state. Furthermore, the squeezing
operator $g^{-a^{\dagger}a}$(or the final squeezing $S(\xi/g^{2})$
in the normalized form) helps to pull apart the two peaks of the cat
state. Hence, by properly adjusting the input signal squeezing parameter
$r$, OPA gain $g$, squeezing $\gamma$ of the target SC state, one
can realize the preparation of SOSC state with large amplitude ($\alpha\ge2$)
with high fidelity. 

To demonstrate the capability of our ($m=4$, $n=1$) configuration
for generating large-amplitude SC states, we compare the heralded
state $\vert\Psi\rangle_{sv,4,1}$ with the ideal SOSC state $\vert\Psi_{\pi}\rangle$
. The Wigner functions of the two states are presented in Fig.\ref{fig:2-3}
(a) and (b), respectively, for OPA gain $g=1.5$ and input squeezing
$r=1.0$. As shown in Fig.\ref{fig:2-3}(a) the heralded state exhibits
a deep negative dip at the origin and two well-separated positive
peaks, which are characteristic features of an odd SC state. The ideal
target SOSC state with parameters $\alpha=2.241$ and $\gamma=1.0$
is displayed in Fig.\ref{fig:2-3} (a), and its Wigner function shows
remarkable similarity to that of the heralded state. The fidelity
between the two states is calculated to be $F=0.991$, confirming
the excellent agreement. This high similarity is further corroborated
by the spatial probability density distributions plotted in in Fig.\ref{fig:2-3}(c)
(blue line for the heralded state, yelllow line for the ideal target
state), which closely match each other across the relevant spatial
region.
\begin{flushleft}
\begin{figure*}
\begin{centering}
\includegraphics[width=14cm]{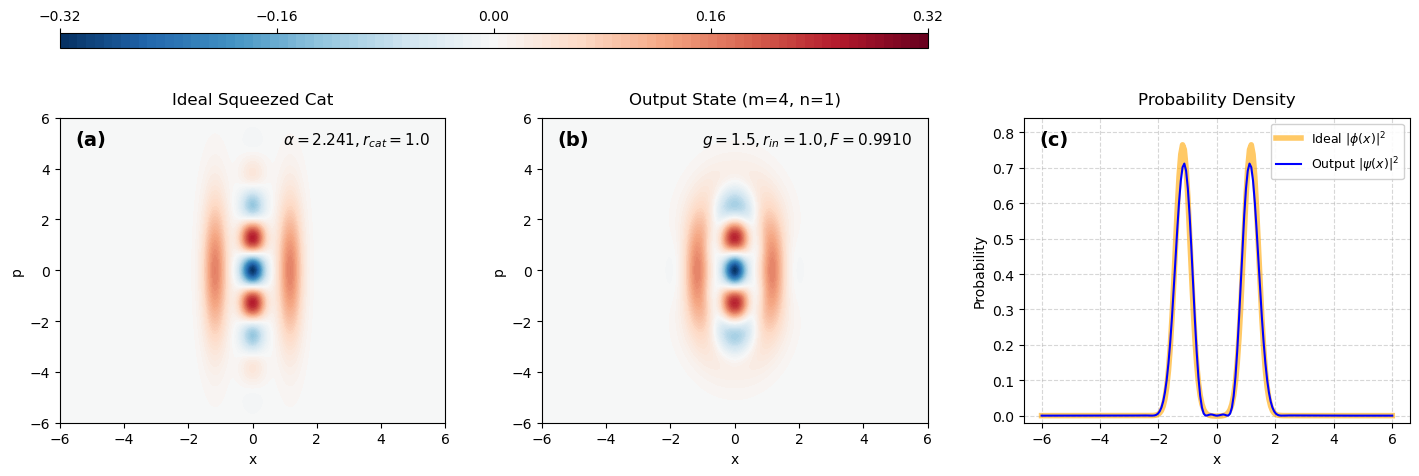}
\par\end{centering}
\caption{\label{fig:2-3}(a) Wigner function of the heralded state $\vert\Psi\rangle_{sv,4,1}$
with $g=1.5$ and $r=1.0$. (b) Wigner function of the ideal SOSC
target with $\alpha=2.241$ and $\gamma=1.0$. The fidelity between
the two states is $F=0.991$. (c) Spatial probability density distributions
of the heralded state (blue curve) and the target state (yellow curve),
confirming the high similarity in real space. The other parameters
are the same as in Fig. \ref{fig:2}.}
\end{figure*}
\par\end{flushleft}

Another significant aspect of the ($m=4$, $n=1$) case is that it
effectively approximates a five-photon subtraction from the squeezed
vacuum state. For the $k=5$ case, the $a^{k}\vert\xi\rangle$is can
give well approximation to SOSC state $\vert\Psi_{sosc}\rangle$ with
at most $\alpha\sim\sqrt{5}\approx2.236$. Our numerical results,
summarized in Table. \ref{tab:1-1} for fixed OPA gain $g=1.5$, and
various input squeezing parameters $r$, confirm this behavior. For
all considered values of $r$, the protocol consistently yields very
high fidelities (all exceeding $0.99$), and the optimal coherent
amplitudes $\alpha$ are all close to $\sqrt{5}$. Specifically, the
relative error between the optimized $\alpha$ and $\sqrt{5}$ is
below $13.6\%$ across the broad parameter ranges. These results demonstrate
that the ($m=4$, $n=1$) configuration provides a viable pathway
for generating high-fidelity, large-amplitude SC states that are essential
for fault-tolerant CV quantum information processing.

\begin{table}
\centering
\caption{\label{tab:1-1}Optimized parameters and fidelity for the approximation
of $\vert\Psi\rangle_{sv,4,1}$by the ideal SOSC state $\vert\Psi_{\pi}\rangle$
at fixed OPA gain $g=1.5$. For each input squeezing $r$, coherent
amplitude $\alpha$, and the corresponding fidelity $F$ are shown.
All fidelity exceed $0.99$, and the optimal $\alpha$ values are
close to $\sqrt{5}\approx2.236$. }

\begin{tabular*}{8cm}{@{\extracolsep{\fill}}cccc}
\toprule 
$r$ & $\gamma$ & $\alpha$  & $F$\tabularnewline
0.857 & 0.924 & 2.100 & 0.995\tabularnewline
0.864 & 0.948 & 2.121 & 0.994\tabularnewline
0.967 & 0.952 & 2.131 & 0.993\tabularnewline
0.848 & 0.964 & 2.152 & 0.992\tabularnewline
0.926 & 0.995 & 2.224 & 0.993\tabularnewline
\bottomrule
\end{tabular*}
\end{table}

In the above subsections we have explicitly investigated the heralded
states produced by the ($m=1$, $n=2$), ($m=1$, $n=3$), ($m=1$,$n=4$)
and ($m=4$, $n=1$) configurations. Among these, the ($m=1$, $n=2$)
and ($m=4$, $n=1$) configurations effectively realize $3$- and
5- photon subtraction from the input SV, respectively. Although not
detailed in the preceding subsections, we have also examined the ($m=3$,
$n=1$) and ($m=5$, $n=1$) cases; they yield $4$\nobreakdash-
and $6$\nobreakdash-photon subtraction with fidelities $\ge0.987$
and $\ge0.961,$ respectively, over the full parameter ranges of $g$
and $r$ considered ( see Table \ref{tab:9-1} ). In fact, we have
systematically scanned all input--output pairs ($m$, $n$) with
$0\le m,n\le7$ and identified the optimal configurations for each
effective subtraction order $k$.

A central observation is that our protocol can effectively implement
$k$-photon subtraction from the SV, but only for specific ($m$,$n$)
pairs. Not every pair satisfying $m+n=k$ yields a SC state with amplitude
$\alpha\approx\sqrt{k}$. Instead, by systematically optimizing the
OPA gain $g$ and the input squeezing parameter $r$, we have identified
specific configurations where the heralded state approximates a certain
type of squeezed SC state with extremely high fidelity and amplitude
$\alpha\sim\sqrt{k}$. For example, the optimized ($m=3$, $n=1$)
and ($m=4$, $n=1$) configurations realize $4$- and $5$-photon
subtraction, respectively, yielding SESC and SOSC states with amplitudes
$\alpha\approx\sqrt{4}$ and $\alpha\approx\sqrt{5}$. In contrast,
the ($m=1$, $n=3$) and ($m=1$, $n=4$) configurations, despite
having $m+n=4$ and $5$, give rise to qualitatively different states:
even\nobreakdash-parity Fock superpositions ( $c_{0}\vert0\rangle+c_{2}\vert2\rangle+c_{4}\vert4\rangle$)
and odd\nobreakdash-parity Fock superpositions ($c_{1}\vert1\rangle+c_{3}\vert3\rangle+c_{5}\vert5\rangle$),
rather than two\nobreakdash-component SC states. This parity\nobreakdash-dependent
selection rule, which has no counterpart in conventional beam splitter
based photon subtraction schemes, represents a new degree of control
in OPA\nobreakdash-based state engineering.

\subsection{Quantum catalysis with $m=n$ }

In the above subsections we focused primarily on configurations with
$m\ne n$, which effectively implement high-order photon subtraction
from the SV with some parity-selected configurations. A distinct and
equally interesting situation occurs when the number of input idler
photons equals the number of detected photons, i.e., $m=n$. In this
case the idler mode starts and ends in the same Fock state $\vert n\rangle$,
the auxiliary mode is therefore restored after the interaction, acting
as a quantum catalyst that enables a nontrivial transformation of
the signal state without being consumed. 

Setting $m=n$ in the general expression Eq. (\ref{eq:9-2}) and using
the properties of the squeezing operator, we obtain a compact form
for the heralded signal state: 
\begin{equation}
\vert\Psi\rangle_{sv,n,n}=R_{n}(a^{\dagger}a)S\left(\frac{\xi}{g^{2}}\right)\vert0\rangle,\ \ R_{n}(x)=\sum^{n}_{k=0}C_{k}x^{k},\label{eq:52-1}
\end{equation}
 where coefficients $C_{k}=\alpha(\xi,g)H(k,n,n)g^{k}$. The state
is therefore a superposition of photon-number-operator powers $(a^{\dagger}a)^{k}$
acting on a SV with scaled squeezing parameter $\xi/g^{2}$. Due to
parity conservation, it contains only even Fock states (see the diagonal
panels in Fig. \ref{fig:4}). In this process, the auxiliary Fock
state $\vert n\rangle$ is not changed by the successful heralding
event, which is the defining feature of photon catalysis \citep{Birrittella:18,ktc9-9rjb}. 

For $n=1$ this reduces to the single-photon catalysis studied in
our previous work \citep{rkzg-sdxn}, where the output state approximates
a small-amplitude SESC state. For $n=2$, we obtain a new family of
non-Gaussian states: 
\begin{align}
\vert\Psi\rangle_{sv,2.2} & \propto\left(C_{0}+C_{1}a^{\dagger}a+C_{2}(a^{\dagger}a)^{2}\right)S\left(\xi/g^{2}\right)\vert0\rangle\nonumber \\
 & =S\left(\xi/g^{2}\right)\left[D_{0}\vert0\rangle+D_{2}\vert2\rangle+D_{4}\vert4\rangle\right],\label{eq:53}
\end{align}
where the explicit expressions of coefficients $D_{0}$, $D_{2}$
and $D_{4}$ can be obtained by straightforward algebra. This state
contains only even Fock components and features a quadratic photon-number
term ($a^{\dagger}a$), which is qualitatively different from states
generated by conventional photon addition or subtraction. By tuning
the OPA gain $g$ and the input squeezing $r$ , the relative weights
$D_{0}:D_{2}:D_{4}$ can be varied continuously. Preliminary analysis
shows that for appropriate parameters the $n=2$ catalyzed state can
offering a new resource for quantum metrology (see Fig. \ref{fig:11-1})
and bosonic error correction. 

\begin{figure}
\includegraphics[width=8cm]{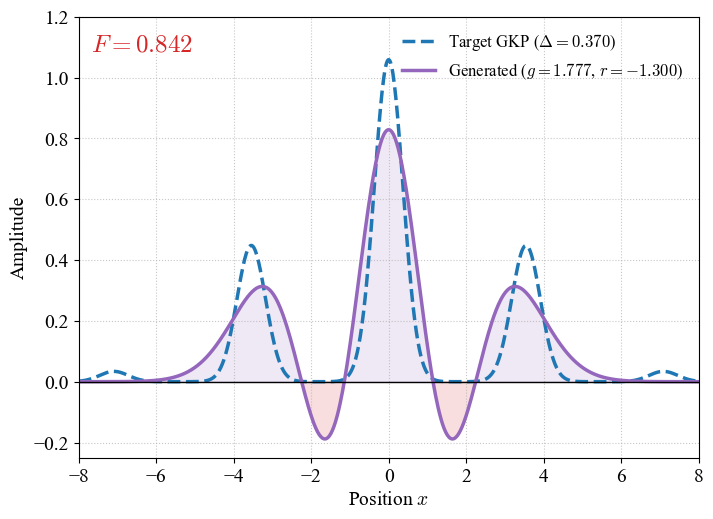}

\caption{\label{fig:10-1}The wave functions of the target GKP state $\psi_{\tilde{0}}(x)$
(blue long dashed curve) with $\triangle=.37$ and our heralded state
$\vert\Psi\rangle_{sv,2,2}$ (solid purple curve) with $g=1.777$,
$r=-1.300.$ The fidelity between two states is $0.842$. The other
parameters are the same as in Fig. \ref{fig:2}.}

\end{figure}

Our state $\vert\Psi\rangle_{sv,2,2}$ can approximate a GKP state.
GKP states encode qubits in CV modes and can protect against small
quadrature shifts in phase space \citep{PhysRevA.64.012310}. Ideal
GKP states are superpositions of infinitely SV states and are unphysical
due to infinite energy requirements. The preparation of optical GKP
states, despite numerous theoretical and experimental efforts {[}62--66{]},
continues to be a formidable task. In practice, finitely squeezed
approximate GKP states are used. The logical basis states $\vert\tilde{0}\rangle$
and $\vert\tilde{1}\rangle$ in the position representation are given
by \citep{PhysRevA.64.012310} 
\begin{align}
\psi_{\tilde{0}}(x) & =\frac{N_{0}}{\left(\pi\triangle^{2}\right)^{1/4}}\sum^{+\infty}_{s=-\infty}e^{-2\pi\triangle^{2}s^{2}-\frac{\left(x-2s\sqrt{\pi}\right)^{2}}{2\triangle^{2}}},\\
\psi_{\tilde{1}}(x) & =\frac{N_{1}}{\left(\pi\triangle^{2}\right)^{1/4}}\sum^{+\infty}_{s=-\infty}e^{-\frac{1}{2}\pi\triangle^{2}\left(2s+1\right)^{2}-\frac{\left[x-(2s+1)\sqrt{\pi}\right]^{2}}{2\triangle^{2}}},
\end{align}
 where $\triangle=e^{-r}$ characterizes the squeezing (standard deviation
) and $N_{0,1}$ are normalization factors. 

Both $\psi_{\tilde{0}}(x)$ and $\psi_{\tilde{1}}(x)$ are even function,
matching the parity of our heralded state $\vert\Psi\rangle_{sv,2,2}$.
As shown in Fig. \ref{fig:10-1}, by optimizing the OPA gain $g$
and input squeezing parameter $r$, we obtain a high fidelity of $0.842$
between the ideal GKP state $\vert\tilde{0}\rangle$ and our state
$\vert\Psi\rangle_{sv,2,2}$ for $g=1.777$, $r=-1.300$, and $\triangle=0.37$
(corresponding to $8.64$ dB squeezing). For this $\triangle=0.37$,
the error probability of the approximate code word is on the order
of $10^{-4}$, which is acceptable for many quantum information tasks.
The generation probability of our heralded state $\vert\Psi\rangle_{sv,2,2}$
for a successful single trial (see Sect. \ref{sec:6} B) is $2.34\times10^{-4}$
with above system parameters. 

Our OPA-based heralding scheme unifies with quantum catalysis: for
$m=n$ it enables higher-order catalysis (beyond single-photon level),
producing squeezed low-dimensional Fock superpositions that are promising
for low-photon metrology and GKP-state approximation \citep{PhysRevA.100.052301}.
This establishes the OPA as a versatile platform for non-Gaussian
state engineering.

\section{\label{sec:5}Phase estimation with our heralded states }

\begin{figure}
\includegraphics[width=8cm]{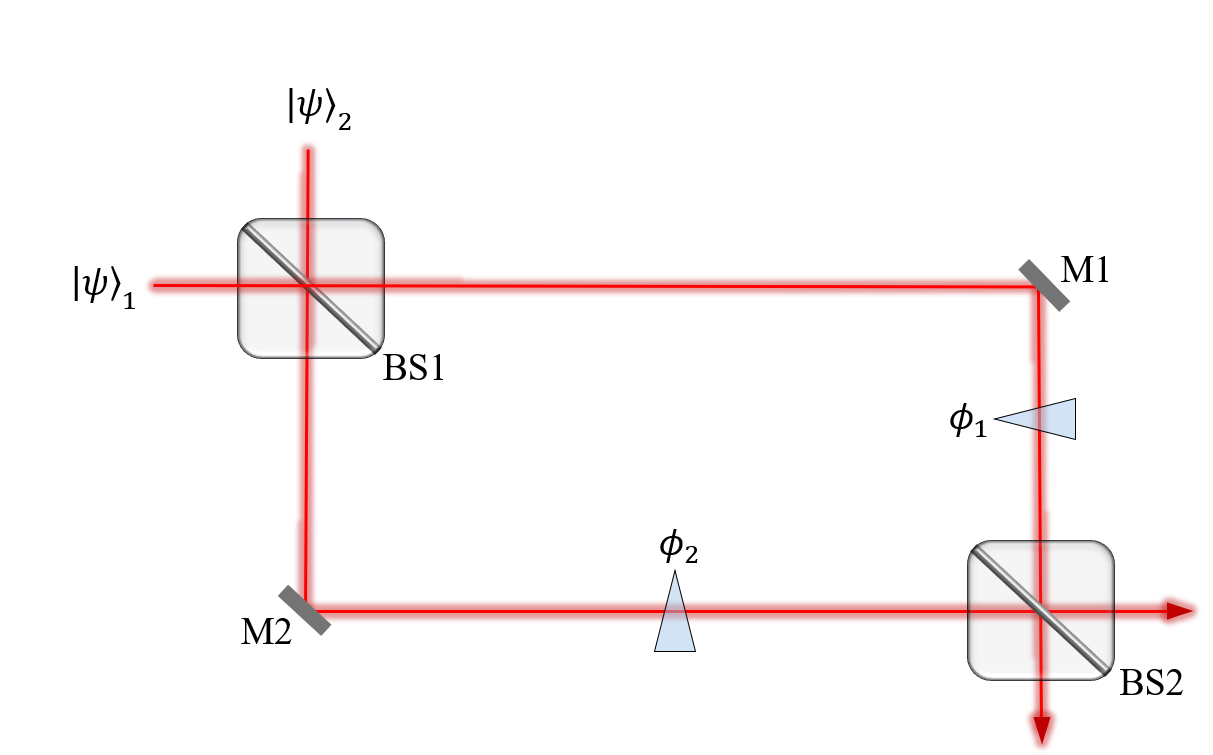}

\caption{\label{fig:10}Mach--Zehnder interferometer setup for estimating
an unknown phase difference. A heralded state, prepared according
to our protocol, is sent into the interferometer. An unknown relative
phase $\phi_{1}-\phi_{2}$ is introduced by two linear phase shifters
located in the two arms. This phase is imprinted onto the initial
quantum state, and subsequent measurements at the output ports allow
the phase to be estimated.}

\end{figure}

As mentioned above, the heralded non-Gaussian states produced by our
protocol have potential applications in CV quantum information processing.
As a concrete example, we demonstrate their advantages in phase estimation.
The schematic of a Mach--Zehnder interferometer (MZI) for phase estimation
is shown in Fig. \ref{fig:10}. Two copies of our heralded state $\vert\Psi\rangle_{sv,n,m}$
are sent into the two inputs ports of the MZI, represented by annihilation
(creation) operators $a$ ($a^{\dagger}$) and $b$ ($b^{\dagger}$)
for the two arms, respectively. These operators obey the bosonic commutation
relations $\left[a,a^{\dagger}\right]=\left[b,b^{\dagger}\right]=1$,
with all other commutators vanishing. The two phase shifters placed
in the arms of the interferometer introduce an unknown relative phase
shift $\phi=\phi_{1}-\phi_{2}$. The unitary evolution imprinted on
the input state is $U(\phi)=\exp\left(i\phi J\right)$, with the phase
generator $J=(a^{\dagger}a-b^{\dagger}b)/2$, i.e., half the photon-number
difference between the two arms. The encoded phase information is
then read out via measurements at the output ports. 

In quantum metrology, the sensitivity of estimating an unknown phase
is characterized by quantum Fisher information (QFI). For a pure quantum
state $\vert\Psi\rangle_{0}=\vert\psi_{1}\rangle\otimes\vert\psi_{2}\rangle$,
the QFI for estimating $\phi$ is defined as 
\begin{equation}
\mathcal{F}=4\left[\langle\partial_{\phi}\Psi_{0}\vert\partial_{\phi}\Psi_{0}\rangle-\vert\langle\Psi_{0}\vert\partial_{\phi}\Psi_{0}\rangle\vert^{2}\right].\label{eq:52}
\end{equation}
 For an unbiased estimation, the phase uncertainty is bound by the
\text{quantum Cram\'{e}r--Rao bound}
\begin{equation}
\triangle^{2}\phi\ge\frac{1}{N_{1}\mathcal{F}},
\end{equation}
where $N_{1}$ is the total number of measurements and in the following
content we take $N_{1}=1$. A higher QFI means a better estimation
precision. For the unitary evolution $U(\phi)=\exp\left(i\phi J\right)$
applied to the input state $\vert\Psi_{0}\rangle$, the QFI reduces
to \citep{2015SR,https://doi.org/10.1002/qute.202100080}
\begin{equation}
\mathcal{F}=4\langle\Psi_{0}\vert\triangle^{2}\mathcal{H}\vert\Psi_{0}\rangle,
\end{equation}
 where $\mathcal{H}=i\left(\partial_{\phi}U^{\dagger}\right)U$ and
$\triangle^{2}\mathcal{H}=(\mathcal{H}-\langle\mathcal{H}\rangle)^{2}$.
For a pure, path-symmetric input state $\vert\Psi_{0}\rangle=\vert\psi_{1}\rangle\otimes\vert\psi_{1}\rangle$
(two identical copies of the heralded state entering the two ports)
with equal mean photon numbers in the two arms, the QFI simplifies
to \citep{PhysRevA.91.013808,PhysRevA.93.033859}
\begin{equation}
\mathcal{F}=2\triangle^{2}(n_{a}),\label{eq:55}
\end{equation}
where $n_{a}=a^{\dagger}a$. This shows that the QFI is determined
solely by the photon\nobreakdash-number variance of the heralded
state.

Previous studies have shown that using coherent and NOON states as
input resources \citep{PhysRevLett.85.2733}, the standard quantum
limit the standard quantum limit (SQL) $\propto$$N$ and Heisenberg
limit (HL) $\text{\ensuremath{\propto}}N^{2}$ can be reached, where
$N$ is the total average photon number of the two arms. For a SV
sate with squeezing parameter $r$, the QFI is $\mathcal{F}_{sv}=N^{2}+2N$,
which can exceed the HL \citep{PhysRevA.91.013808}. We now examine
whether our heralded states generated from SV via the OPA can outperform
the SV itself. In this work, we define the HL limit as $\mathcal{F}_{HL}=N^{2}$
(the scaling achieved by NOON states).

\begin{figure}
\includegraphics[width=8cm]{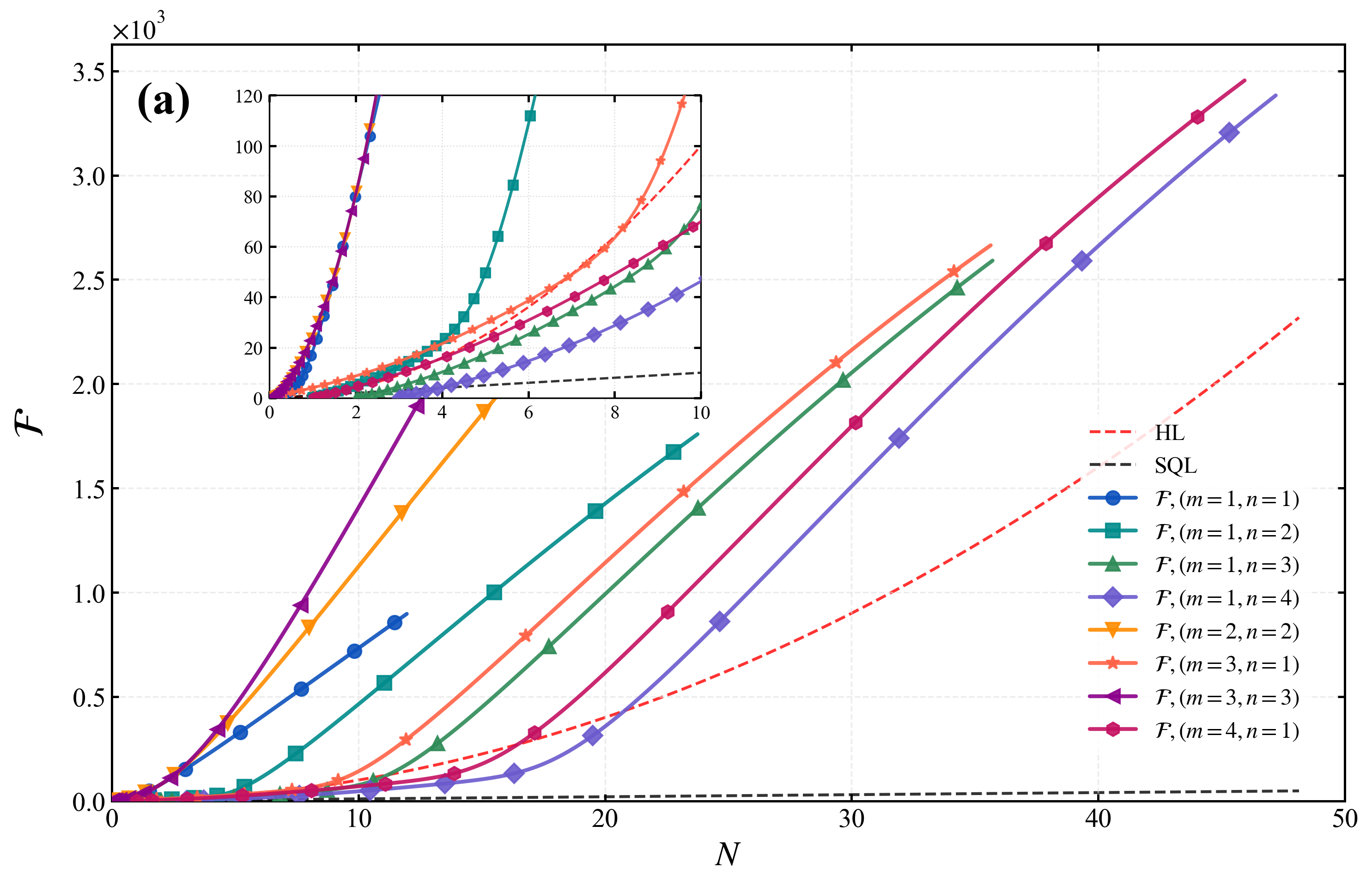}

\includegraphics[width=8cm]{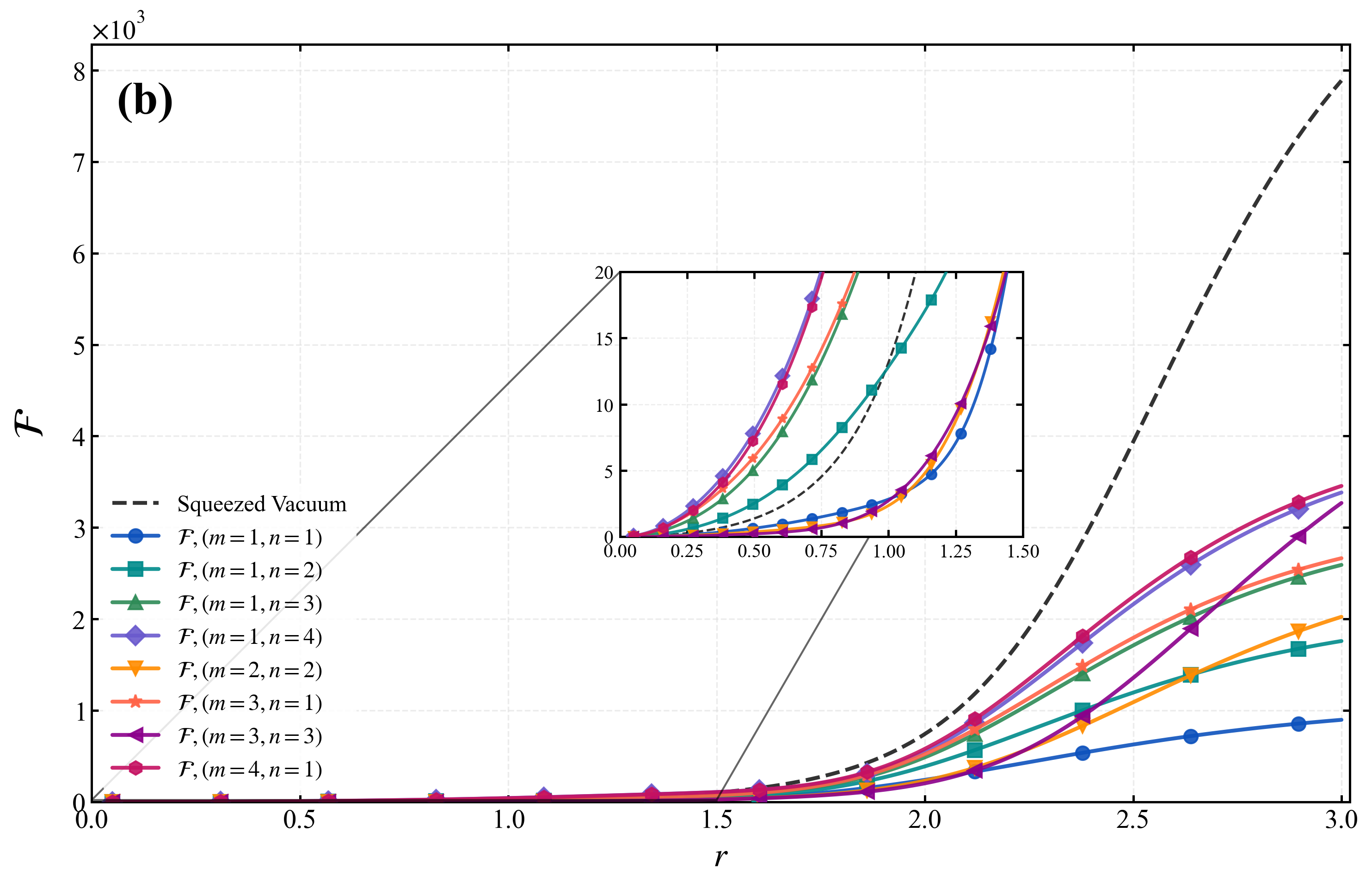}

\caption{\label{fig:11-1}Quantum Fisher information (QFI) $\mathcal{F}$ plotted
as a function of the average photon number $N$ and squeezing parameter
in (a) and (b), respectively, with different photon heralded sates
corresponding to some configurations ($m$, $n$ ). Here, SQL and
HL denote the standard quantum limit (SQL) $N$ and Heisenberg limit
(HL) $N^{2}$, and the OPA gain fixed to $g=1.04$. The other parameters
are the same as in Fig. \ref{fig:2}.}
\end{figure}

Based on Eq. (\ref{eq:55}), we can evaluate the QFI analytically
for the heralded output states $\vert\Psi\rangle_{sv,m,n}$ by different
configurations ($m$, $n$). As specific examples, we choose eight
configurations: ($m=1$, $n=1$) to ($1$, $4$), ($m=3$, $n=1$),
($m=2$, $n=2$), ($m=3$, $n=3$) and ($m=4$, $n=1$). The numerical
results are shown in Fig. \ref{fig:11-1}. 

In Fig. \ref{fig:11-1} (a), we plot the QFI of thses states as a
function of the squeezing parameter (ranging from $0$ to $3$), which
determines the average photon number. The OPA gain is fixed to $g=1.04$.
For large photon number, the QFI values of the heralded states are
all higher than the SQL and the HL, especially for the ($m=1$, $n=4$)
, ($m=4$, $n=1$), ($m=1,$$n=3$) and ($m=3,$$n=1$) configurations.
In the low-photon-number regimes, however, the QFI of the above configurations
(except ($m=1$, $n=1$) , ($m=1$, $n=2$), ($m=3,$$n=1$)) are
lower than the HL, while (except ($m=1$, $n=1$) , ($m=1$, $n=2$),
($m=3,$$n=1$)) exceed the HL, as shown in the inset of Fig. \ref{fig:11-1}
(a). Despite their limited performance at large photon numbers, the
catalyzed states with $m=n$ achieve super-Heisenberg scaling in the
low-photon-number regime, which is particularly relevant for loss-sensitive
or low-flux metrology. The degradation below the Heisenberg limit
at higher photon numbers is caused by the intrinsic Fock cutoff at
$2n$ photons, which bounds the mean photon number and saturates the
photon-number variance.

In Fig. \ref{fig:11-1} (b), we present the QFI as a function of the
squeezing parameter $r$ for the same configurations, again with $g=1.04$,
and compare them with the SV reference. In weak squeezing input regime
(see the inset), all configurations except ($m=1$, $n=1$) yield
a better phase estimation performance than the SV. In the moderate\nobreakdash-squeezing
regime, the QFI of ($m=1$, $n=4$) , ($m=4$, $n=1$) and ($m=3,$$n=1$)
remains larger than that of the SV. However, in the strong\nobreakdash-squeezing
regime, the scaling of the QFI for most configurations approaches
the SQL but remains inferior to the HL. 

One can understand the mechanism of enhanced phase estimation with
our heralded states as follows. For our pure and separable initial
states, the QFI can be expressed in terms of the Mandel $\mathcal{Q}=\left[\triangle^{2}(n_{a})-\langle n_{a}\rangle\right]/\langle n_{a}\rangle$
factor as 
\begin{equation}
\mathcal{F}=N\left(1+\mathcal{Q}\right).\label{eq:56}
\end{equation}
Our calculations show that for the heralded states considered, the
Mandel $\mathcal{Q}$ factor is approximately proportional to the
average photon number. Consequently, the proportionality constant,
which is larger than unity, allows the QFI to numerically exceed the
HL in the large-photon-number limit, while still maintaining the same
quadratic scaling.

From the above discussion we can conclude that the phase estimation
precision using our heralded states can achieve excellent performance
and even surpass the HL over a wide range of input squeezing parameters,
provided the OPA gain $g$ is chosen appropriately. Among the states
considered, the ($m=1$, $n=4$) , ($m=4$, $n=1$) and ($m=1,$$n=3$)
configurations, which possess high complexity and negativity, give
the best performance. Furthermore, for phase estimation in the weak\nobreakdash-squeezing
regime, the heralded states produced by our protocol are preferable
to the SV itself.

\section{\label{sec:6}Practical considerations }

In this section, we analyze the experimental feasibility of our heralded
non-Gaussian state generation scheme. We consider two key practical
aspects: the detrimental effects of photon loss and dephasing, and
the overall success probabilities for different parameter configurations,
including the use of high-repetition-rate laser sources to overcome
low single-trial success probabilities.

\subsection{\label{subsec:A}Photon loss and dephasing}

In any realistic optical implementation, quantum states are subject
to decoherence due to interaction with the environment. For bosonic
modes, the dominant imperfections are single-photon loss and dephasing.
The dynamics of the system under these noise channels can be described
by the master equation
\begin{equation}
\frac{d\rho(t)}{dt}=\left(\kappa\mathcal{D}\left[a\right]\rho+\kappa_{\phi}\mathcal{D}\left[a^{\dagger}a\right])\right)\rho(t)\label{eq:38}
\end{equation}
 where $\kappa$ and $\kappa_{\phi}$ are the photon loss and dephasing
rates, respectively, and $\mathcal{D}\left[A\right]\rho\equiv2A\rho A^{\dagger}-A^{\dagger}A\rho-\rho A^{\dagger}A$
is the Lindblad dissipator. Note that loss and dephasing errors are
generated by the jump operator $a$ and $a^{\dagger}a$ respectively.
We say that a system state is loss-dominated (dephasing-dominated)
if $\kappa\gg\kappa_{\phi}$($\kappa_{\phi}\gg\kappa$). 

To assess the robustness of our scheme against environmental decoherence,
we numerically simulate the effects of single-photon loss and dephasing
on the fidelity of the heralded states. The dynamics is governed by
the master equation in Eq. (\ref{eq:38}). For the configurations
($m=1$, $n=2$) and ($m=4$, $n=1$), which produce approximated
SOSC states with amplitude $\alpha\sim1.73$ and $\alpha\sim2.237$,
respectively, we evaluate the fidelity with respect to the ideal target
state $\vert\Psi_{sosc}\rangle$, with fixed squeezing $r=1.0$ and
OPA gain $g=1.5$. 

\begin{figure}
\includegraphics[width=8cm]{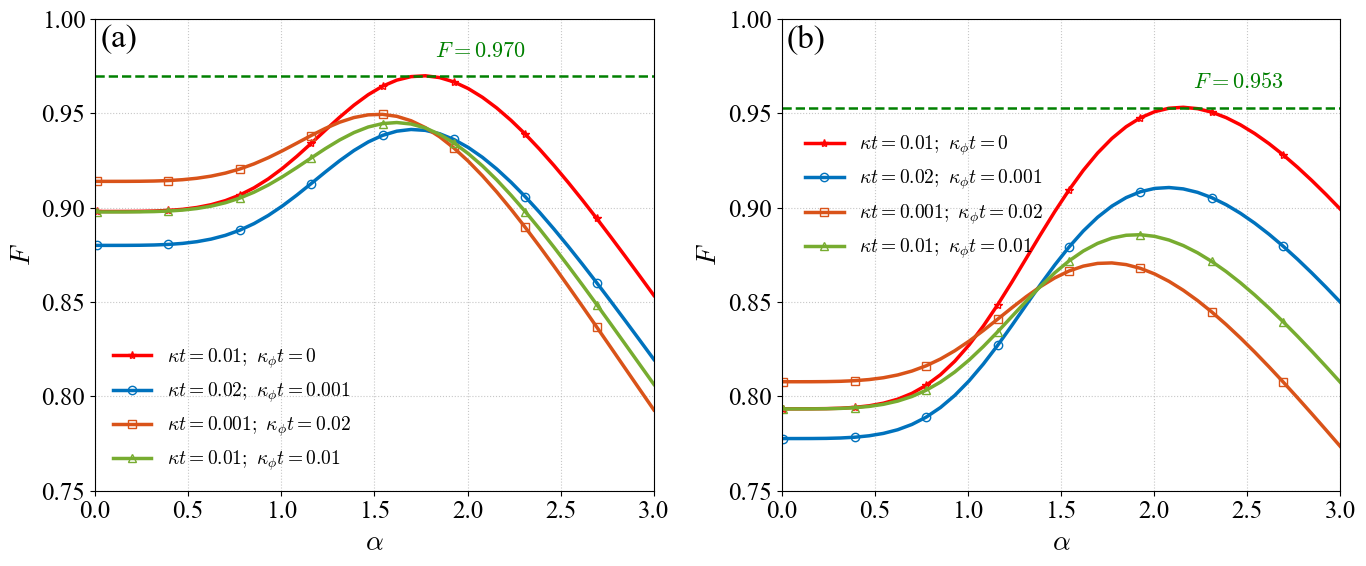}

\caption{\label{fig:11}Fidelity of heralded states $\vert\Psi\rangle_{sv,1,2}$(a)
and $\vert\Psi\rangle_{sv,4,1}$ (b) with respect to the ideal SOSC
state $\vert\Psi_{sosc}\rangle$ under photon loss and dephasing.
OPA gain and input SV's squeezing are set to $g=1.5$ and $r=1.0$.
The other parameters are the same as in Fig. \ref{fig:2}.}
\end{figure}

Figure \ref{fig:11} presents the optimal fidelity as a function of
the target coherent amplitude $\alpha$ for different noise scenarios:
pure photon loss ($\kappa_{\phi}=0$), loss\nobreakdash-dominated
case ($\kappa/\kappa_{\phi}=20$), dephasing\nobreakdash-dominated
case ( $\kappa_{\phi}/\kappa=20$) and equivalent rates ($\kappa=\kappa_{\phi}$).
The fidelity of ($m=4$, $n=1$) configuration more sensitive to losses
than that of the ($m=1$, $n=2$) case. For the pure photon loss with
$\kappa t=0.01$, the optimal overlap remains above $0.95$ in both
configurations, with optimal target parameters $\alpha=1.84$, $\gamma=0.703$
for ($m=1$, $n=2$) and $\alpha=2.16$, $\gamma=0.966$ for ($m=4$,
$n=1$) ( see blue solid curves in Fig.\ref{fig:11} (a) and (b),
respectively. )

Comparing the photon-losss dominated ($\kappa/\kappa_{\phi}=20$)
and dephasing- dominated ($\kappa_{\phi}/\kappa=20$) cases reveals
different effects. For ($m=1$, $n=2$) configuration, the optimal
fidelity in dephasing-domintated case is higher than that photon loss-dominated
case, while the former also yields a larger coherent amplitude $\alpha$
(see blue and orange curves in Fig.\ref{fig:11} (a) ). In contrast,
for ($m=4$, $n=1$) configuration, both the optical fidelity and
the corresponding coherent amplitude of target SOSC state are lower
in the dephasing-dominated than in the photon loss-dominated case.
Furthermore, we observe that the equal amounts of photon loss and
dephasing ($\kappa=\kappa_{\phi}$) give better performance than either
the photon loss-dominated or dephasing-dominated case for both configurations.
As we can see, dephasing significantly diminishes the interference
fringes of the quantum state in the phase space, leading to a decrease
in the optimal fidelity between our heralded state and target states. 

\begin{figure}
\includegraphics[width=8cm]{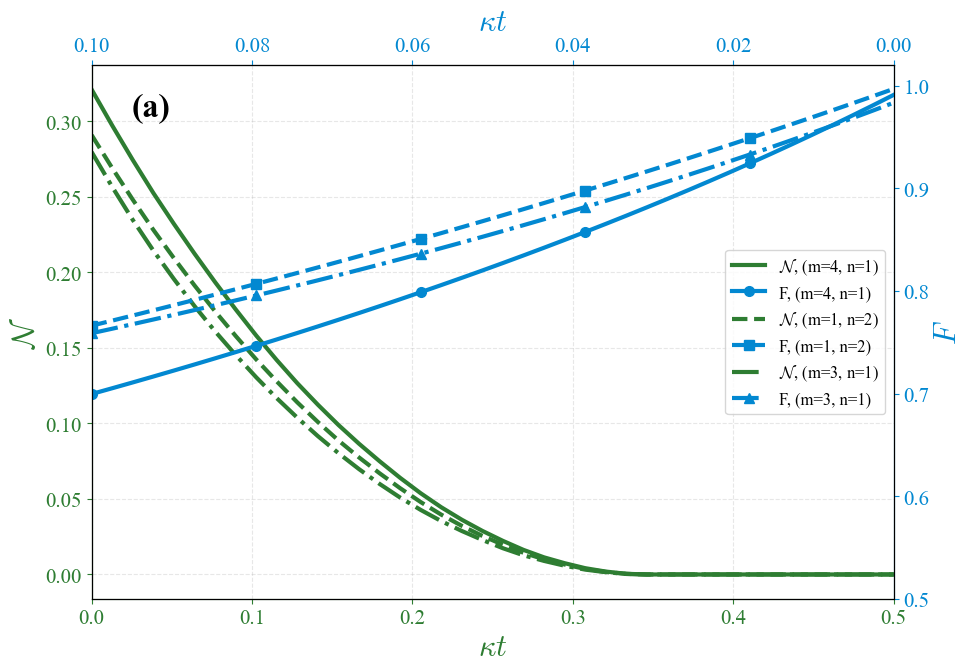}

\includegraphics[width=8cm]{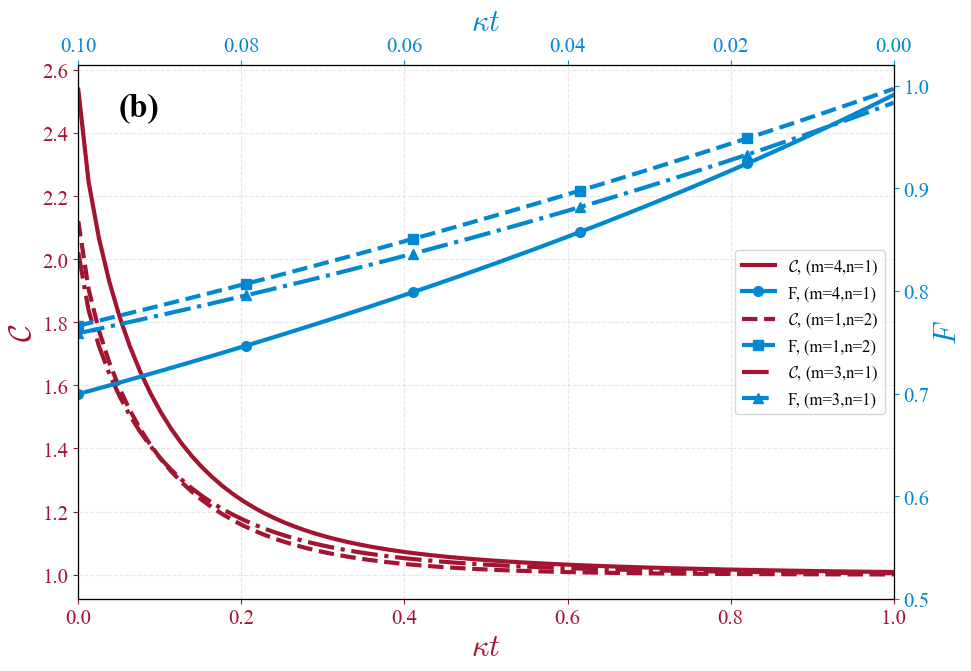}

\caption{\label{fig:13}Changes of the fidelity $F$, negativity $\mathcal{N}$
and complexity $\mathcal{C}$ of our heralded states $\vert\Psi\rangle_{sv,m,n}$
with ($m=1$, $n=2$), ($m=3$, $n=1$) and ($m=4$, $n=1$) configurations
vs photon loss rate $\kappa t$. Here, the most optimized fidelity
for different configurations as follow: (i) ($m=1$, $n=2$) configuration,
$\alpha_{opt}=1.795$, $\gamma_{opt}=0.706$, $F=0.997$; (ii) ($m=3$,
$n=1$) configuration, $\alpha_{opt}=1.9$, $\gamma_{opt}=0.979$,
$F=0.984$; (iii) ($m=4$, $n=1$) configuration, $\alpha_{opt}=2.2$
, $\gamma_{opt}=0.986$, $F=0.991$. The other parameters are the
same as in Fig. \ref{fig:2}.}
\end{figure}

To further illustrate the effects of pure photon loss on the optimal
fidelity between our heralded states--which effectively realize three-,
four- and five-photon subtraction--and on different types of squeezed
SC states, we show in Fig. \ref{fig:13} the changes of the most optimized
fidelity $F$, negativity $N$ and complexity $\mathcal{C}$ of our
heralded states $\vert\Psi\rangle_{sv,m,n}$ for ($m=1$, $n=2$),
($m=3$, $n=1$) and ($m=4$, $n=1$) as functions of the photon loss
rate $\kappa t$. 

As can be seen, the optimal fidelity $F$ for all three configurations
decreases with increasing $kt$. The most optimized fidelities--
$F=0.997$ (for ($1$, $2$) with $\alpha_{opt}=1.795$, $\gamma_{opt}=0.706$),
$F=0.984$ (for ($3$, $1$) with $\alpha_{opt}=1.9$, $\gamma_{opt}=0.979$),
and $F=0.991$ (for ($4$, $1$) with $\alpha_{opt}=2.2$ , $\gamma_{opt}=0.986$)--remain
high ($F\ge0.9$) in the regimes $\kappa t\le0.038$, $\kappa t\le0.031$
and $\kappa t\le0.025$, respectively. These high fidelities drop
from $\ge0.98$ at zero loss to approximately $0.766$, $0.759$ and
$0.700$ at $\kappa t=0.1$ (i.e., $10\%$ single-photon loss). 

The negativity $\mathcal{N}$ and complexity $\mathcal{C}$ of our
heralded states also decreases with increasing $\kappa t$, eventually
reaching their minimum values $0$ and $1$, respectively. Interestingly,
even at $10\%$ single photon loss ($\kappa t=0.1$), the negativity
and complexity of these states are not small. For $\kappa t\gtrapprox0.33$,
the negativity $\mathcal{N}$ becomes zero while the complexity $\mathcal{C}$
remains larger than $1$. For our heralded states, $\mathcal{C}$
does not become exactly $1$ even when the photon loss rate is as
high as $\kappa t=1$ (see Fig. \ref{fig:13} (a) and (b)). This indicates
that the negativity $\mathcal{N}$ and complexity $\mathcal{C}$ of
our heralded states have excellent robustness against the single-photon
loss, especially for those states that initially possess high complexity
and negativity. 

Overall, these results demonstrate that our protocol is compatible
with state\nobreakdash-of\nobreakdash-the\nobreakdash-art experimental
conditions: moderate single\nobreakdash-photon loss or dephasing
alone preserves high fidelities, confirming the practical feasibility
of the scheme. Nevertheless, when photon loss and dephasing are present
at comparable rates, the fidelity can be significantly reduced, while
negativity $\mathcal{N}$ and complexity $\mathcal{C}$ remain with
considerable values. This resilience indicates that our states retain
useful quantum resources under realistic loss, with potential advantages
in fault\nobreakdash-tolerant quantum computing \citep{R2023,PhysRevResearch.3.033275,hr5f-lvy7},
lossy quantum metrology \citep{PhysRevA.93.033859,PhysRevA.96.062304},
and state certification \citep{PRXQuantum.2.020333}.

\subsection{\label{subsec:B}Success probability}

The total probability per trial for successfully generating a heralded
non\nobreakdash-Gaussian state in our scheme can be expressed as
\begin{equation}
P_{trial}=\eta_{det}\times P_{Fock}(m,\tau)\times P_{sv}(r,g,m,n)\label{eq:39-1}
\end{equation}
where $P_{Fock}(m,g^{\prime})$ is the probability to generate $m$-photon
Fock state at the idler input via spontaneous parametric down\nobreakdash-conversion
(SPDC) using a pump with gain parameter $g^{\prime}$; $P_{sv}(r,r,g,n)$
is the conditional heralding probability for the OPA stage, i.e.,
the probability that the idler output contains exactly $n$-photons
given an input $m$-photon Fock state and a SV (squeezing parameter
$r$ ) in the signal mode; $\eta_{det}$ is the system detection efficiency
of the PNRD, which includes intrinsic quantum efficiency, coupling
losses, and filtering losses.

The $m$-photon Fock state can be prepared by seeding an SPDC source
(OPA) with vacuum and heralding on the detection of $m$ photons.
The SPDC process is described by the same two\nobreakdash-mode squeezing
unitary as the OPA {[}see Eq.\,(\ref{eq:1}{]}, so that the $m$-photon
state can be expressed as 
\begin{equation}
\vert m\rangle_{Fock}={}_{b}\langle m\vert\left(\mathbb{I}\otimes\Pi_{n}\right)S(\tau^{\prime})\left(\vert0\rangle_{a}\otimes\vert0\rangle_{b}\right).\label{eq:40}
\end{equation}
The preparation probability $P_{Fock}(m,g^{\prime})$ is given by
the norm square of this unnormalized state. For the parameters used
in this work, we take $g^{\prime}=1.0483$ and compute the relevant
probabilities. Modern superconducting nanowire single\nobreakdash-photon
detectors can be operated in a photon\nobreakdash-number\nobreakdash-resolving
mode with high efficiency \citep{2020K}. We adopt a conservative
value $\eta_{det}=0.9$ ($90\%$) which is routinely achieved in state\nobreakdash-of\nobreakdash-the\nobreakdash-art
experiments \citep{2012,Zhang:24}.

Table IX lists the success probabilities for various ($m$, $n$ )
combinations, with input squeezing $r=1.0$($\approx8.69$dB) and
OPA gain $g=1.5$. $P_{Fock}$ and $P_{sv}$ are obtained from numerical
calculations ; $P_{trial}$ is computed as their product times $\eta_{det}$.
\\

\begin{table}
\caption{\label{tab:9}Success probabilities of our heralded state with ($m$,
$n$). Parameters: $r=1.0$, $g^{\prime}=1.0483$, $g=1.5$ and $\eta_{det}=0.9$. }

\begin{tabular}{c c l l l}
\toprule 
\multicolumn{2}{c}{($m$, $n$)} & $P_{Fock}(m,g^{\prime})$ & $P_{sv}(r,g,m,n)$ & $P_{trial}$\tabularnewline
\midrule
\multicolumn{2}{c}{($1$, $1$)} & \multirow{4}{2cm}{$8.19\times10^{-2}$} & $1.58\times10^{-1}$ & $1.16\times10^{-2}$\tabularnewline
\multicolumn{2}{c}{($1$, $2$)} &  & $1.56\times10^{-1}$ & $1.15\times10^{-2}$\tabularnewline
\multicolumn{2}{c}{($1$, $3$)} &  & $1.29\times10^{-1}$ & $9.57\times10^{-3}$\tabularnewline
\multicolumn{2}{l}{($1$, $4$)} &  & $1.00\times10^{-1}$ & $7.42\times10^{-3}$\tabularnewline
\multicolumn{2}{l}{($2$, $2$)} & $7.38\times10^{-3}$ & $8.11\times10^{-2}$ & $5.38\times10^{-4}$\tabularnewline
\multicolumn{2}{l}{($3$, $1$)} & $6.64\times10^{-4}$ & $3.95\times10^{-2}$ & $2.36\times10^{-5}$\tabularnewline
\multicolumn{2}{c}{($4$, $1$)} & $5.98\times10^{-5}$ & $1.15\times10^{-2}$ & $6.20\times10^{-7}$\tabularnewline
\multicolumn{2}{c}{($5$, $1$)} & \multirow{2}{2cm}{$5.38\times10^{-6}$} & $1.15\times10^{-2}$ & $5.58\times10^{-8}$\tabularnewline
\multicolumn{2}{c}{($5$, $2$)} &  & $6.17\times10^{-3}$ & $2.99\times10^{-8}$\tabularnewline
\bottomrule
\end{tabular}
\end{table}

For the ( $m=1$, $n=2$ ) configuration, which produces a high\nobreakdash-fidelity
SOSC state with amplitude $\alpha\approx1.73$, the total trial probability
is $P_{trial}\approx1.4\times10^{-2}=1.4\%$. This is remarkably high
for a heralded non\nobreakdash-Gaussian state source and is directly
comparable to the success rates reported in state\nobreakdash-of\nobreakdash-the\nobreakdash-art
GPS experiments \citep{Takase:22,PhysRevA.103.013710}. The dominant
contribution comes from $P_{sv}=1.56\times10^{-1}$, which is favorable
because $n=2$ detection is efficient and $m=1$ (single photon) can
be generated near\nobreakdash-deterministically.

For the ($m=4$, $n=1$ ) configuration (amplitude $\alpha\sim\sqrt{5}\approx2.236$),
$P_{trial}\approx6.20\times10^{-7}$. This low value is primarily
due to the small $P_{Fock}(4,\tau)=5.98\times10^{-5}$, reflecting
the difficulty of generating a four\nobreakdash-photon Fock state
in a single trial with SPDC process. This obstacle could, in principle,
be overcome by employing alternative, near-deterministic methods for
preparing higher-order Fock states, such as those demonstrated in
cavity and circuit QED systems \citep{PhysRevLett.125.093603,PhysRevResearch.2.033489,PhysRevA.110.042421}.
In particular, the conditional measurement protocol of Zhang \& Jing
\citep{PhysRevA.110.042421}, which is conceptually similar to our
heralding approach, can generate Fock states $\vert m\rangle$ with
fidelity exceeding $99\%$ in less than 30 measurement cycles, even
for $n\sim10$. Integrating such a high-fidelity, efficient source
of $m$-photon Fock states with our OPA scheme would effectively eliminate
the $P_{Fock}$ factor, potentially boosting $P_{trial}$ by several
orders of magnitude. Nevertheless, with current SPDC technology, low
single\nobreakdash-trial probabilities may still raise concerns about
experimental feasibility; it is comparable to or even higher than
the heralding probabilities reported in successful multiphoton state
engineering experiments. For instance, in the experimental preparation
of three\nobreakdash-photon\nobreakdash-added coherent states, Fadrn�
et al. achieved heralding probabilities on the order of $10^{-9}$
for similar multiphoton operations \citep{89}.

Low single\nobreakdash-trial probabilities can be effectively overcome
by operating the pump laser at a high repetition rate. A standard
80\,MHz repetition rate yields $R_{herald}=P_{trial}\times f_{rep}=6.20\times80\times10^{6}\approx50Hz$
(50 states per second), permitting accumulation of millions of heralded
states per day, sufficient for high\nobreakdash-fidelity tomography
and quantum information tasks. State\nobreakdash-of\nobreakdash-the\nobreakdash-art
laser systems can achieve repetition rates in the GHz regime (e.g.,
$1$\,GHz to $50$\,GHz) \citep{Jornod:23,Wakui:20,Zhang:24} even
in terahertz (THz) \citep{Zhang:24} regimes, the rate can be boosted
to thousands of hertz. For the ($m=1$, $n=2$ ) case, under the $1.55$-\textgreek{\textmu}m-driven
difference-frequency generation with a $32$ MHz repetition rate which
can operated in the regime of OPA \citep{Liu:23}, it can give approximately
$4.76\times10^{5}$ state per second, which is exceptionally high
and exceeds the rates reported in recent experimental demonstrations
of SC state generation \citep{chen2023}. 

Thus, despite varying single\nobreakdash-trial probabilities, the
combination of heralded operation, high\nobreakdash-repetition\nobreakdash-rate
pumping, and efficient PNRDs \citep{2023} makes all configurations
experimentally feasible. Our OPA\nobreakdash-based scheme is well
within reach of current optical quantum technologies.

\section{\label{sec:7}Conclusion}

In this study, we have proposed and analyzed a multiphoton heralding
scheme based on an OPA that generates non\nobreakdash-Gaussian states
of light with controllable higher\nobreakdash-Fock components from
a SV input. In contrast to our previous work \citep{rkzg-sdxn}, which
considered only the ($1$,1) case, here we uncover a parity\nobreakdash-dependent
selection rule that determines whether a heralded state approximates
a SC state or a multi\nobreakdash-Fock superposition. Such a selection
rule does not arise in conventional $k$\nobreakdash-photon subtraction
schemes based on a BS. In our OPA scheme, it emerges naturally and
provides an additional degree of control. By varying the input--output
photon pair ($m$, $n$), the OPA gain $g$, and the input squeezing
$r$, one can steer the output state between SC\nobreakdash-like
and multi\nobreakdash-Fock superpositions. We have found that systematic
optimization of the $g$ and $r$ yields a certain type of squeezed
SC state with extremely high fidelity and amplitude at most $\alpha\approx\sqrt{k}$
for some specific configurations satisfying $m+n=k$. Table\, \ref{tab:9-1}
summarizes the optimized configurations that effectively realize $k$-photon
subtraction for $k=1$ to $7$, along with the optimized corresponding
target parameters (squeezing parameter $\gamma$, coherent amplitude
$\alpha$, and fidelity $F$) and the type of non\nobreakdash-Gaussian
state produced. 

Beyond state engineering, we have systematically characterized the
non\nobreakdash-classicality and structural richness of our heralded
states using Wigner negativity $\mathcal{N}$ and complexity $\mathcal{C}$.
The results show that states with higher $\mathcal{N}$ and $\mathcal{C}$
(e.g., ($m=1$, $n=4$) and ($m=4$, $n=1$)) exhibit superior performance
in phase estimation, achieving QFI that surpasses both the SQL and
the HL (defined as $N^{2}$ ) over a wide range of parameters. Moreover,
these states demonstrate remarkable robustness against photon loss:
even when $\mathcal{N}$ vanishes under strong loss, $\mathcal{C}$
persists, highlighting the resilience of their structural complexity.

Regarding experimental feasibility, the total success probability
for the high\nobreakdash-performance ($m=1$, $n=2$) configuration
reaches $1.15\%$ per trial, while the large\nobreakdash-amplitude
($m=4$, $n=1$) configuration has a single\nobreakdash-trial probability
on the order of $10^{-7}$. These probabilities are comparable to
those of other heralded non\nobreakdash-Gaussian state schemes (e.g.,
mutli-photon\nobreakdash-added coherent states \citep{89} and cluster
states \citep{PhysRevLett.97.110501}). Although low at first glance,
such probabilities can be effectively overcome by operating the pump
laser at a high repetition rate transforming per\nobreakdash-trial
probabilities into absolute heralding rates. This places the predicted
performance of our protocol on par with state\nobreakdash-of\nobreakdash-the\nobreakdash-art
experimental demonstrations. Nevertheless, practical implementation
demands high phase stability in the idler interferometer, low photon
loss, and high detector efficiency to maintain output fidelity. With
continued progress in high\nobreakdash-repetition\nobreakdash-rate
lasers, low\nobreakdash-loss integrated photonics, and high\nobreakdash-efficiency
superconducting detectors, our scheme provides a scalable and viable
pathway for advanced quantum state engineering.

Our protocol thus provides a flexible and integrated platform for
engineering higher\nobreakdash-order photon subtraction and generating
a rich family of non\nobreakdash-Gaussian states---ranging from
squeezed SC states to even\nobreakdash- and odd\nobreakdash-parity
Fock superpositions. In particular, the catalyzed state with $m=n=2$
is shown to approximate a finite-energy GKP codeword with a fidelity
of 0.842 (see Fig. \ref{fig:10-1}), demonstrating the potential of
our scheme for generating bosonic error-correction resources. This
work establishes the OPA as a versatile, reconfigurable tool for non\nobreakdash-Gaussian
state engineering, and opens new directions for exploring the interplay
between structural complexity, quantum metrology, and fault\nobreakdash-tolerant
quantum information processing.

\begin{table*}[htbp]
\centering
\caption{\label{tab:9-1}The realization of effective $k$-photon (up to $7$-photon)
subtraction with our protocol.}

% 使用最纯粹的居中对齐 c，彻底消除排版逻辑冲突
\begin{tabular}{c c c c c c c c c c}
\toprule 
\multicolumn{2}{c}{$k$} & ($m$, $n$) & SC type & $\kappa_{\phi}$ & $\alpha$ & \multicolumn{2}{c}{$F$} & \multicolumn{2}{c}{{\footnotesize Realization with BS scheme (Exps)}}\tabularnewline
\midrule 
\multicolumn{2}{c}{$1$} & ($1$, $0$) or ($0$, $1$) & odd & $0.30\pm0.06$ & $0.9\pm0.1$ & \multicolumn{2}{c}{$0.998\pm0.001$} & \multicolumn{2}{c}{Ref. \citep{2006,Takase:22}}\tabularnewline
\multicolumn{2}{c}{$2$} & ($1$, $1$) & even & $0.59\pm0.07$ & $1.4\pm0.1$ & \multicolumn{2}{c}{$0.997\pm0.001$} & \multicolumn{2}{c}{Ref. \citep{PhysRevLett.101.233605}}\tabularnewline
\multicolumn{2}{c}{$3$} & ($1$, $2$) & odd & $0.45\pm0.15$ & $1.7\pm0.1$ & \multicolumn{2}{c}{$0.998\pm0.001$} & \multicolumn{2}{c}{Ref. \citep{PhysRevA.82.031802,Endo:23}}\tabularnewline
\multicolumn{2}{c}{$4$} & ($3$, $1$) & even & $1.07\pm0.07$ & $2.0\pm0.1$ & \multicolumn{2}{c}{$0.970\pm0.010$} & \multicolumn{2}{c}{Ref. \citep{endo2025}}\tabularnewline
\multicolumn{2}{c}{$5$} & ($4$, $1$) & odd & $1.15\pm0.15$ & $2.2\pm0.1$ & \multicolumn{2}{c}{$0.993\pm0.002$} & \multicolumn{2}{c}{--\footnote{No experiment reported to date}}\tabularnewline
\multicolumn{2}{c}{$6$} & ($5,$ $1$) & even & $1.24\pm0.04$ & $2.45\pm0.1$ & \multicolumn{2}{c}{$0.920\pm0.010$} & \multicolumn{2}{c}{--}\tabularnewline
\multicolumn{2}{c}{$7$} & ($5,$ $2$) & odd & $0.95\pm0.04$ & $2.6\pm0.1$ & \multicolumn{2}{c}{$0.979\pm0.005$} & \multicolumn{2}{c}{--}\tabularnewline
\bottomrule
\end{tabular}
\end{table*}

\begin{acknowledgments}
This study was supported by the National Natural Science Foundation
of China (No. 12365005). 
\end{acknowledgments}

\bibliographystyle{apsrev4-1}
\bibliography{Reference}

\end{document}